\documentclass[aps,dvips]{revtex4}
\usepackage{epsfig}	% Include figure files
\usepackage{bm}		% bold math

\newcommand{\dd}{\textrm{d}}

\begin{document}
\title{Inhomogeneous hard-sphere mixtures: Manifestations of structural crossover}
\author{C.~Grodon}
\affiliation{Max-Planck-Institut f{\"u}r Metallforschung,
Heisenbergstra{\ss}e 3, D-70569 Stuttgart, Germany\\}
\affiliation{Institut f{\"u}r Theoretische und Angewandte Physik,
Universit{\"a}t Stuttgart, Pfaffenwaldring 57, D-70569 Stuttgart, Germany\\}
\author{M.~Dijkstra}
\affiliation{Debye Institute, Soft Condensed Matter Group, Utrecht University,
Princetonplein 5, 3584 CC Utrecht, The Netherlands\\}
\author{R.~Evans}
\affiliation{H.H. Wills Physics Laboratory, University of Bristol, Bristol
BS8 1TL, UK\\}
\affiliation{Max-Planck-Institut f{\"u}r Metallforschung,
Heisenbergstra{\ss}e 3, D-70569 Stuttgart, Germany\\}
\affiliation{Institut f{\"u}r Theoretische und Angewandte Physik,
Universit{\"a}t Stuttgart, Pfaffenwaldring 57, D-70569 Stuttgart, Germany\\}
\author{R.~Roth}
\affiliation{Max-Planck-Institut f{\"u}r Metallforschung,
Heisenbergstra{\ss}e 3, D-70569 Stuttgart, Germany\\}
\affiliation{Institut f{\"u}r Theoretische und Angewandte Physik,
Universit{\"a}t Stuttgart, Pfaffenwaldring 57, D-70569 Stuttgart, Germany\\}
%\author{C. Grodon$^1$$^\ast$\thanks{$^\ast$ Corresponding author. E-mail: grodon@mf.mpg.de}, M. Dijkstra$^2$, R. Evans$^{3,1}$ and R. Roth$^1$
%\affil{$^1$ Max-Planck-Institut f{\"u}r Metallforschung, Heisenbergstra{\ss}e 3, D-70569 Stuttgart, Germany,\\
%and Institut f{\"u}r Theoretische und Angewandte Physik, Universit{\"a}t Stuttgart, Pfaffenwaldring 57, D-70569 Stuttgart, Germany}
%\affil{$^2$ Debye Institute, Soft Condensed Matter Group, Utrecht University, Princetonplein 5, 3584 CC Utrecht, The Netherlands}
%\affil{$^3$ H.H. Wills Physics Laboratory, University of Bristol, Bristol BS8 1TL, UK}
%}

%%%%%%%%%%%%%%%%%%%%%%%%%%%%%%%%%%%%%%%%%%%%%%%%%%%%%%%%%%%%%%%%%%%%%%%%%%%%%%%%
%%%%%%%%%%%%%%%%%%%%%%%%%%%%%%%%%%%%%%%%%%%%%%%%%%%%%%%%%%%%%%%%%%%%%%%%%%%%%%%%
%%%%%%%%%%%%%%%%%%%%%%%%%%%%%%%%%%%%%%%%%%%%%%%%%%%%%%%%%%%%%%%%%%%%%%%%%%%%%%%%
\begin{abstract}
We study various manifestations of structural crossover in the properties of a binary mixture of hard-spheres. For homogeneous mixtures that are sufficiently asymmetric, there is a crossover line in the phase diagram such that for thermodynamic states that lie on one side, the decay of the three bulk pair correlation functions is oscillatory with a common wavelength approximately equal to the diameter of the smaller spheres, and for states on the other side, the common wavelength is approximately the diameter of the bigger spheres. Using density functional theory we show that structural crossover manifests itself in the {\em intermediate} range decay of i) the one-body density profiles of a hard-sphere mixture adsorbed at planar walls, ii) the effective (depletion) potential between two big hard-spheres immersed in the same binary mixture and iii) the solvation force, or excess pressure, of the same mixture confined between two planar hard-walls. We determine exactly the structural crossover line for a one-dimensional binary mixture of hard-rods and present evidence, based on density functional theory calculations and Monte-Carlo simulations, for structural crossover in homogeneous binary mixtures of hard-disks in two dimensions. By considering a multicomponent mixture of hard-spheres, with an appropriate bimodal distribution of diameters, we argue that structural crossover should still occur in the presence of polydispersity and that our results could be relevant to colloidal mixtures where correlation functions can be measured using real-space techniques.
\end{abstract}

\maketitle

%%%%%%%%%%%%%%%%%%%%%%%%%%%%%%%%%%%%%%%%%%%%%%%%%%%%%%%%%%%%%%%%%%%%%%%%%%%%%%%%
%%%%%%%%%%%%%%%%%%%%%%%%%%%%%%%%%%%%%%%%%%%%%%%%%%%%%%%%%%%%%%%%%%%%%%%%%%%%%%%%
%%%%%%%%%%%%%%%%%%%%%%%%%%%%%%%%%%%%%%%%%%%%%%%%%%%%%%%%%%%%%%%%%%%%%%%%%%%%%%%%
\section{Introduction} 

In 1969 Michael Fisher and Ben Widom \cite{bib:fisher69} published an elegant paper in which they conjectured that there should be a line in the phase diagram of a simple one-component fluid at which the character of the longest range decay of the total pair correlation function $h(r)$ changes from monotonic to exponentially damped oscillatory. Their conjecture, which was based on exact results for certain one-dimensional models, pertains to fluids which exhibit liquid-gas phase separation.  Although the Fisher--Widom (FW) line constitutes a sharp, well-defined crossover line, in say the temperature-density plane, it does not imply any accompanying non-analyticity in the free energy of the bulk fluid and therefore no phase transition.  The significance of the FW line for the structure of wall-fluid and liquid-gas interfaces and for the asymptotic decay of the solvation force in a confined fluid was not appreciated until much later \cite{bib:evans94,bib:evans93}.  Indeed the first attempt at calculating the FW line using a realistic theory of liquids (for a square-well model) was not reported until 1993 \cite{bib:evans93}.  Subsequently there have been several investigations of FW crossover for various types of fluids using a variety of liquid state theories and the first computer simulation determination of the FW line (for a truncated Lennard-Jones fluid) was reported in 2000 \cite{bib:dijkstra00}.  In the present paper we follow the spirit of FW in that we seek qualitative changes in the asymptotic decay of correlation functions corresponding to different regions of the phase diagram.  Unlike FW, we focus on binary mixtures and investigate changes in the wavelength of the oscillations in various structural quantities; we are not concerned with crossover from monotonic to oscillatory decay rather with crossover from asymptotic oscillatory decay with one particular wavelength to decay with a different wavelength.

In a recent paper \cite{bib:grodon04}, denoted I, we showed that for sufficiently asymmetric binary mixtures of (additive) hard-spheres there is a sharp structural crossover line in the $(\eta_b, \eta_s)$ phase diagram at which the common wavelength of the longest range oscillations in the three bulk pair correlation functions $h_{bb}(r)$, $h_{ss}(r)$ and $h_{bs}(r)$ changes discontinuously from approximately $\sigma_b$, the diameter of the bigger species, to $\sigma_s$, the diameter of the smaller species.  $\eta_b$ and $\eta_s$ denote the packing fractions of big and small spheres.  We also showed, using density functional theory (DFT) and Monte Carlo simulation (MC), that structural crossover manifests itself at {\em intermediate} range, not just at longest range.  It follows that crossover might be observed in real-space measurements of the pair correlation functions in colloidal mixtures.

Here we investigate what repercussions this new crossover line might have for the structure of {\em inhomogeneous} hard-sphere mixtures.  We also examine other properties of mixtures seeking further manifestations of structural crossover.

Our paper is arranged as follows.  In subsection \ref{sec:theorybulk} we summarise the main features of the general theory of the asymptotics of pair correlation functions in binary mixtures and outline the nature of structural crossover found in bulk fluids in paper I.  Subsection \ref{sec:theorywall} presents DFT results for one-body density profiles, $\rho_i(z)$, $i=s,b$, of the mixture at two types of planar wall.  These demonstrate clearly that structural crossover is just as easy to observe at wall-fluid interfaces as in bulk.  Indeed a single pole approximation for $\rho_i(z)$ is just as effective as the corresponding approximation \cite{bib:grodon04} for the pair correlation functions $h_{ij}(r)$.  In subsection \ref{sec:depletion} we consider the effective depletion $W(r)$ potential between two big hard spheres immersed in the binary mixture.  Using a particle insertion method, based on Widom's \cite{bib:widom63} potential distribution theorem, combined with DFT we show that the intermediate and long range behaviour of $W(r)$ provides a clear signature of crossover, i.e.~the wavelength of the oscillations in the depletion potential exhibits the same variation with the thermodynamic state point $(\eta_b, \eta_s)$ as is found in $h_{ij}(r)$.  An equivalent conclusion is reached in subsection \ref{sec:solvationforce} for the solvation force $f_s(L)$ which arises when the binary hard-sphere mixture is confined between two planar hard walls separated by a distance $L$.  The oscillations in this {\em thermodynamic} quantity mimic closely those found in $h_{ij}(r)$ and in the wall-fluid density profiles $\rho_i(z)$.  Throughout we focus on a mixture with size ratio $q=\sigma_s/\sigma_b=0.5$. However, there is nothing special about that particular choice for $q$ as we pointed out in paper I.

In section \ref{sec:polydispersity} we extend our analysis to the case of polydisperse hard-sphere mixtures, bearing in mind that polydispersity is always relevant in colloidal systems.  We calculate, using DFT, the density profiles for a multicomponent mixture with a large, but finite, number of species adsorbed at a hard-wall. The size distributions are chosen to mimic a bimodal system with two maxima that are very weakly overlapping.  In spite of the maxima being very broad we find a clear signature of crossover in the density profile of the species whose diameter is equal to $\sigma_b^{max}$, the position of the maximum in the size distribution of big spheres.  We return to homogeneous binary mixtures in section \ref{sec:rods} and investigate structural crossover in low dimensional fluids.  For one-dimensional hard-rod mixtures there are exact results for the pair direct correlation functions which allow us to determine exactly the relevant poles of the Fourier transforms of $h_{ij}(r)$.  We find crossover, demonstrating that this phenomenon arises in an exactly solvable model. The crossover line has a similar shape to that found for the hard-sphere mixture with the same size ratio.

Two-dimensional hard-disk mixtures are of particular interest from an experimental point of view. Using video microscopy one can measure accurately correlation functions in colloidal suspensions that are confined by light fields to two dimensions. This technique might provide a means of testing the crossover scenario experimentally. We perform DFT calculations and Monte-Carlo simulations to verify that structural crossover also occurs for hard-disk mixtures in dimension $d=2$.

We conclude, in section \ref{sec:discussion}, with a discussion of our results and their possible relevance for other mixtures and for experiment.

%%%%%%%%%%%%%%%%%%%%%%%%%%%%%%%%%%%%%%%%%%%%%%%%%%%%%%%%%%%%%%%%%%%%%%%%%%%%%%%%
%%%%%%%%%%%%%%%%%%%%%%%%%%%%%%%%%%%%%%%%%%%%%%%%%%%%%%%%%%%%%%%%%%%%%%%%%%%%%%%%
%%%%%%%%%%%%%%%%%%%%%%%%%%%%%%%%%%%%%%%%%%%%%%%%%%%%%%%%%%%%%%%%%%%%%%%%%%%%%%%%
\section{Asymptotic Decay of Correlation Functions} 

In this section we recall some of the predictions of the general theory of the asymptotics of correlation functions laid out in Refs.~\cite{bib:evans94,bib:evans93}; a summary of the main results is given in Ref.~\cite{bib:evans96} which adopts a density functional theory perspective from the outset. We consider a binary mixture of big ($b$) and small ($s$) particles with bulk (reservoir) densities $\rho_i$ and chemical potentials $\mu_i$, $i=s,b$. Numerical results will be presented for hard-sphere mixtures.

%%%%%%%%%%%%%%%%%%%%%%%%%%%%%%%%%%%%%%%%%%%%%%%%%%%%%%%%%%%%%%%%%%%%%%%%%%%%%%%%
%%%%%%%%%%%%%%%%%%%%%%%%%%%%%%%%%%%%%%%%%%%%%%%%%%%%%%%%%%%%%%%%%%%%%%%%%%%%%%%%
\subsection{Pair Correlations in Bulk Mixtures}
\label{sec:theorybulk}

The nature of structural crossover of the total correlation functions $h_{ij}(r) = g_{ij}(r) - 1$, where $g_{ij}(r)$ is the radial distribution function, was studied in detail for homogeneous binary hard-sphere mixtures in paper I. The analysis is based on calculating the leading-order poles of $\hat h_{ij}(k)$, the Fourier transform of $h_{ij}(r)$.

In the bulk mixture $h_{ij}(r)$ are related to the pair direct correlation functions $c_{ij}^{(2)}(r)$ via the Ornstein-Zernike (OZ) equations
\begin{equation}
h_{ij}(r_{12}) = c_{ij}^{(2)}(r_{12}) + \sum_{k=s,b} \rho_k \int \dd^3 r_3 c_{ik}^{(2)}(r_{13}) h_{kj}(r_{32}),
\end{equation} 
where $r_{ij}=|\mathbf{r}_i-\mathbf{r}_j|$. 

If we suppose that the direct correlation functions are known the OZ equations can be solved easily in Fourier space. $\hat h_{ij}(k)$ can be written as 
\begin{equation}
\hat h_{ij}(k) = \frac{\hat N_{ij}(k)}{\hat D(k)},
\label{eqn:FourierOZ}
\end{equation}
where the denominator common to all three equations is given by
\begin{equation}
\hat D(k) = [1-\rho_s \hat c_{ss}^{(2)}(k)][1-\rho_b \hat c_{bb}^{(2)}(k)] - \rho_s\rho_b \hat c_{bs}^{(2)}(k)^2.
\label{eqn:denominator}
\end{equation}
$\hat c_{ij}^{(2)}(k)$ is the Fourier transform of $c_{ij}^{(2)}(r)$.

\begin{figure}
\centering\epsfig{file=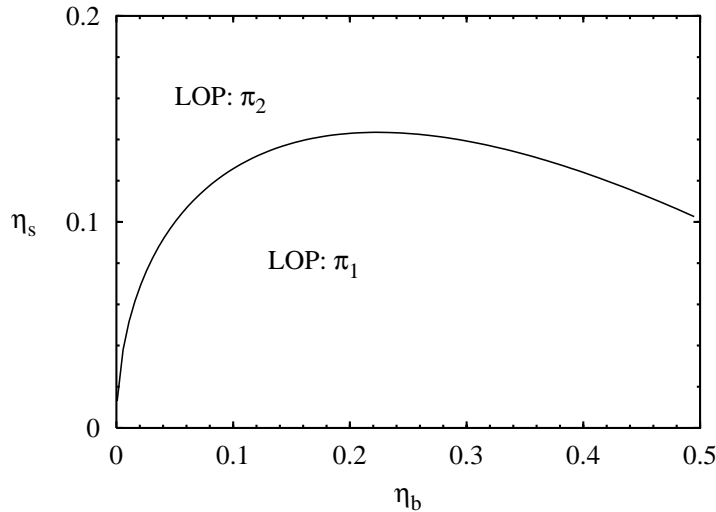,width=10cm}
\vspace{0.5cm}
\caption{\label{fig:crossoverline} Crossover line for a binary hard-sphere mixture with size ratio $q=0.5$. On each side of the line the asymptotic behaviour is dominated by different leading order poles (LOP) labelled $\pi_1$ and $\pi_2$. For small values of the packing fraction $\eta_s$, $\pi_1$ dominates and the wavelength of the oscillations in $h_{ij}(r)$ is $\sim \sigma_b$. On increasing $\eta_s$ at fixed $\eta_b$ the wavelength decreases gradually until crossover occurs. Then $\pi_2$ dominates and the oscillations have a wavelength $\sim 0.5 \sigma_b$. Generally at the crossover line the wavelength of the longest ranged oscillations changes discontinuously by a factor of about $q$. These results were obtained using the Percus-Yevick approximation for the direct correlation functions $\hat c_{ij}^{(2)}(k)$ --- see I. Note that the crossover line begins at the origin and we have truncated it at $\eta_b = 0.5$.} 
\end{figure}

From the formal solution (\ref{eqn:FourierOZ}) one obtains an expression for the total correlation functions $h_{ij}(r)$  in real space by performing the inverse Fourier transform via the residue theorem. We assume that the singularities of $\hat h_{ij}(k)$ are (simple) poles \cite{bib:evans93}. In order to determine the asymptotic behaviour we must consider the leading order pole (LOP) contribution to $r h_{ij}(r)= (2 \pi)^{-1}\sum_n R_n^{ij} \exp{(i p_n r)}$, where $p_n$ are the aforementioned poles, given by the zeros of $\hat D(k)$. $R_n^{ij}$ is the residue of $k \hat h_{ij}(k)$ corresponding to the pole $p_n$. Poles are either pure imaginary, $p= i a_0$, or occur as a conjugate complex pair $p=\pm a_1 + i a_0$. The pole, or pair of poles, with the smallest imaginary part $a_0$ determines the slowest exponential decay and is termed the LOP. This pole determines the ultimate, $r \to \infty$, decay of $h_{ij}(r)$. Note that in the case of complex poles all three total correlation functions show a {\em common} oscillatory asymptotic decay
\begin{equation}
r \, h_{ij}(r) \sim A_{ij} \exp{(-a_0 r)} \cos{(a_1 r - \Theta_{ij})}, \quad r \to \infty.
\label{eqn:decaysphere}
\end{equation}
The imaginary and real parts of the LOP determine the characteristic decay length $a_0^{-1}$ and the wavelength of oscillations $2 \pi / a_1$, respectively. These length scales are common for all pairs $ij$, whereas the phases $\Theta_{ij}$ and the amplitudes $A_{ij}$ are dependent on the indices $ij$. As emphasised in I this observation is remarkable when one considers a binary system where the sizes of the two species are quite different. Clearly at high concentrations of the small particles one would expect the wavelength to be determined by the size of the small particles whereas at low concentrations one would expect that the wavelength of oscillations is set by the size of the big particles. Explicit calculations, based on DFT, show that the wavelength of the longest ranged oscillations $2 \pi/a_1$ 
changes discontinuously at some sharp crossover line in the phase diagram of binary hard-sphere mixtures \cite{bib:grodon04}.
In figure \ref{fig:crossoverline} the crossover line is plotted for a hard-sphere mixture with size ratio $q=\sigma_s/\sigma_b=0.5$. $\sigma_s$ and $\sigma_b$ denote the hard-sphere diameters and results are presented in terms of the packing fractions $\eta_i = \pi \rho_i \sigma_i^3 / 6$, $i=s,b$. $\pi_1$ denotes the pole with the longest wavelength $2 \pi / a_1 \sim \sigma_b$, whereas $\pi_2$ is that with the second longest wavelength $2 \pi / a_1 \sim \sigma_b/2$. There are higher-order poles $\pi_3$, $\pi_4$, etc.~\cite{bib:grodon04} but these are not germane to figure \ref{fig:crossoverline}. The genesis of the crossover line is given in figure 2 of paper I where the trajectories of poles are plotted. Note that close to a crossover point the two poles $\pi_1$ and $\pi_2$ have similar imaginary parts, $a_0$ and $\tilde a_0$, and both contribute to the oscillatory decay of the total correlation functions at intermediate values of $r$, provided that the corresponding amplitudes, $A_{ij}$ and $\tilde A_{ij}$, are similar in size, i.e.~near crossover we expect
\begin{equation}
r \, h_{ij}(r) \sim A_{ij} \exp{(-a_0 r)} \cos{(a_1 r - \Theta_{ij})} + \tilde A_{ij} \exp{(-\tilde a_0 r)} \cos{(\tilde a_1 r - \tilde \Theta_{ij})}
\label{eqn:decaycrossover}
\end{equation}
for large $r$. The first contribution corresponds to $\pi_1$ with $a_1 \sim 2 \pi / \sigma_b$ and the second to $\pi_2$ with $\tilde a_1 \sim 2 \pi / (q \sigma_b)$. Crossover occurs when $a_0=\tilde a_0$. 

In I we investigated the decay of $h_{ij}(r)$ at intermediate range by performing DFT calculations of the density profiles of both species in the presence of a fixed (test) particle exerting an external potential on the other particles in the fluid. Using both the Rosenfeld \cite{bib:rosenfeld89} and White Bear version \cite{bib:roth02} of fundamental measure DFT we found that the crossover behaviour predicted by the asymptotic (pole) analysis manifests itself at intermediate separations as well as at longest range, $r\to \infty$. This is illustrated in figures 6 and 7 of I. Away from the crossover line the result from the LOP approximation (\ref{eqn:decaysphere}) provides an accurate fit to the DFT results except for small separations, $r<2\sigma_b$, where many poles begin to contribute. For state points close to the crossover line the two-pole approximation (\ref{eqn:decaycrossover}) provides as good fit to the DFT results as is achieved by equation (\ref{eqn:decaysphere}) away from the line; the presence of two oscillatory wavelengths implied by  equation (\ref{eqn:decaycrossover}) is manifest in the numerical DFT data. This observation will be important later when we consider manifestations of structural crossover for {\em inhomogeneous} fluids.

\begin{figure}
\centering\epsfig{file=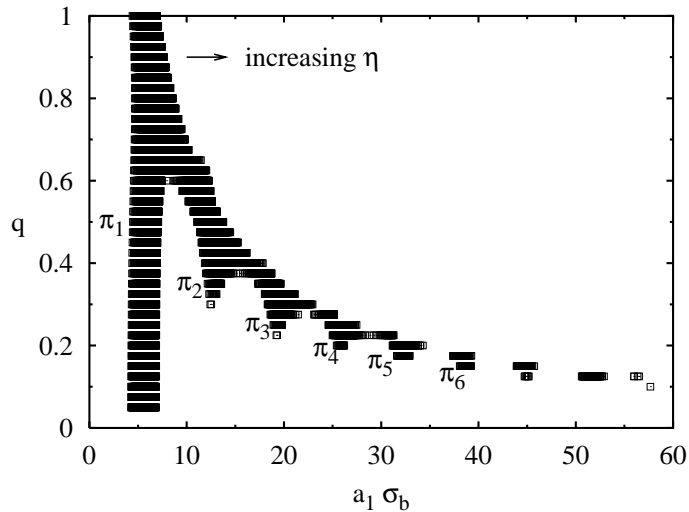,width=10cm}
\vspace{0.5cm}
\caption{\label{fig:fundamentalplot} Real part $a_1$ of the leading order poles calculated for fixed size ratios $q$ and various combinations of packing fractions: generally $\eta=\eta_s+\eta_b$ increases from left to right up to a maximum of 0.5. For the less symmetric mixtures with $q\lesssim 0.65$ separate branches $\pi_1$, $\pi_2$, etc.~arise and there are no leading order poles with values of $a_1$ lying in the gaps. Structural crossover corresponds to a 'jump' from a value of $a_1$ associated with one branch of poles to a value associated with another --- see text.}
\end{figure}

We complete our summary of structural crossover in homogeneous binary hard-sphere mixtures by presenting, in figure \ref{fig:fundamentalplot}, results for the real part, $a_1$, of the leading order poles calculated for a full range of size ratios $q$. As in I, the poles were determined by calculating the zeros of $\hat D(k)$ in equation (\ref{eqn:denominator}) using the Percus-Yevick results for the pair direct correlation functions. For the more symmetric mixtures, $q\gtrsim 0.7$, the LOP is $\pi_1$ for all packing fractions so there is a continuous variation of $a_1$ in this regime; there is no crossover. For more asymmetric mixtures there are discontinuities in $a_1$ and separate branches appear in figure \ref{fig:fundamentalplot}. For example, for $q=0.5$ there is a gap between the branches corresponding to $\pi_1$ and $\pi_2$, and crossover corresponds to jumping between these branches as the packing fractions are varied. As $q$ is reduced further additional branches appear and for $q=0.35$ crossover can occur between $\pi_1$ and $\pi_2$ and between $\pi_2$ and $\pi_3$ or directly between $\pi_1$ and $\pi_3$. Similarly for $q=0.2$ crossover can occur between $\pi_1$ and $\pi_4$, $\pi_4$ and $\pi_5$ or directly between $\pi_1$ and $\pi_5$.

The origin of the crossover lines shown in figure 5 of I can be understood from this 'master' plot of the poles. Note that for small size ratios some of the packing fractions considered in the plot correspond to state points that lie within the fluid-solid coexistence region --- see I.

%%%%%%%%%%%%%%%%%%%%%%%%%%%%%%%%%%%%%%%%%%%%%%%%%%%%%%%%%%%%%%%%%%%%%%%%%%%%%%%%
%%%%%%%%%%%%%%%%%%%%%%%%%%%%%%%%%%%%%%%%%%%%%%%%%%%%%%%%%%%%%%%%%%%%%%%%%%%%%%%%
\subsection{Density Profiles at a Planar Wall}
\label{sec:theorywall}

Earlier treatments \cite{bib:evans94,bib:evans93,bib:evans96} of the asymptotics of correlation functions emphasised that, for fluids with short-ranged interatomic potentials, the pole structure given by the zeros of $\hat D(k)$ in equation (\ref{eqn:denominator}) determines not only the decay of $h_{ij}(r)$ in a bulk mixture but also the decay of the one body density profiles at wall-fluid and liquid-vapour interfaces. Given this observation we should expect to find manifestations of structural crossover in the properties of inhomogeneous mixtures and we seek these in the present subsection.

Perhaps the most direct way of analysing the asymptotic decay, $z\to\infty$,
of the density profiles, $\rho_{wi}(z)$, of a binary mixture adsorbed at a
planar wall is to start from the wall-particle OZ equations, i.e.~we consider
a ternary mixture with species $s,b$ plus a third species whose density
$\rho_3\to 0$. The third component, whose diameter is $\sigma_3$, can be
considered as a wall that exerts an external potential on the mixture of $s$
and $b$ particles. The resulting OZ equations can be expressed as 
\begin{equation}
\hat h_{wi}(k) = \frac{N_{wi}(k)}{\hat D(k)},
\label{eqn:FourierOZwall}
\end{equation}
$i=s,b$, where $\hat h_{wi}(k)$ denotes the Fourier transform of $h_{wi}(r)$, the wall-particle total correlation function and the third species is now labelled $w$ (wall). Note that the form of (\ref{eqn:FourierOZwall}) is valid for any value of $\sigma_3$ and the repercussions for the density profiles hold for a spherical wall, $\sigma_3<\infty$, for which the wall-particle total correlation function $h_{wi}(r)$ has spherical symmetry, as well as for a planar one,  $\sigma_3 \to \infty$, for which we have planar symmetry for the total correlation functions, $h_{wi}(z)$. $z$ is then the distance from the {\em planar} wall and $\hat h_{wi}(k)$ is replaced by $\tilde h_{wi}(k)$, the {\em one-dimensional} Fourier transform. The denominator $\hat D(k)$ is common to both species $i=s$ and $b$ particles and is identical to that appearing in equation (\ref{eqn:denominator}): it depends upon the pair direct correlation functions of the bulk mixture of $s$ and $b$ whose densities $\rho_s$ and $\rho_b$ are fixed by the reservoir far from the wall. In the planar case the numerator $N_{wi}(k)$ can be written in terms of the one-dimensional Fourier transforms of the wall-particle direct correlation functions $c_{wi}(z)$ which depend, of course, upon the wall fluid potentials $V_w^i(z)$. Provided the latter decay faster than correlation functions in the bulk fluid the asymptotic decay of $h_{wi}(z)$ will be determined be the LOPs of $\hat D(k)$. Since bulk correlations decay as $\exp{(-a_0 r)}$, where $a_0$ is the imaginary part of the LOP, this requires $V_w^i(z)$ to decay faster than $\exp{(-a_0 z)}$. In the calculations to be described below we shall consider only hard walls or walls whose attractive potential is of finite support. 

The planar wall-particle density profile is given by $\rho_{wi}(z)\equiv\rho_i[h_{wi}(z)+1]$, $i=s,b$, and from equation (\ref{eqn:FourierOZwall}) it follows that the asymptotic decay is described by
\begin{equation}
\rho_{wi}(z) - \rho_i \sim B_i^w \exp{(-a_0 z)} \cos{(a_1 z - \Theta_i^w)}, \quad z \to\infty
\label{eqn:decaywall2}
\end{equation}
where $\rho_i$ denotes the bulk density of species $i$. Although the decay length $a_0^{-1}$ and wavelength $2\pi/a_1$ are the same for both species, the amplitude $B_i^w$ and phase $\Theta_i^w$ depend on the species label and on the choice of wall-fluid potentials. Clearly equation (\ref{eqn:decaywall2}) is the analogue of equation (\ref{eqn:decaysphere}). Note that the additional factor $1/r$ in the latter arises from taking a three-dimensional rather than a one-dimensional Fourier transform.

Equation~(\ref{eqn:decaywall2}) implies that if structural crossover occurs in the correlation functions of a bulk mixture one should observe equivalent behaviour in the one-body density profiles of the same mixture, near the same crossover state point, adsorbed at a planar wall. We examine this prediction below. 

There are alternative ways of deriving equation (\ref{eqn:decaywall2}) based on i) a DFT approach \cite{bib:evans96}, or ii) analysis of the exact integral equations for the density profiles in planar geometry \cite{bib:henderson92}.

\begin{figure}
\centering\epsfig{file=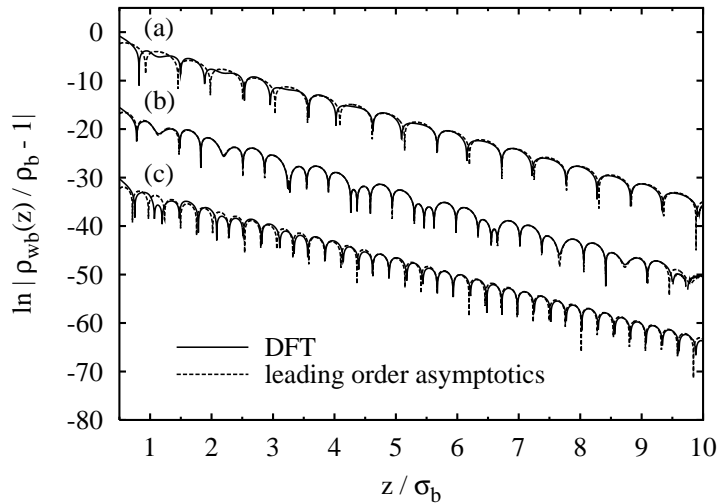,width=10cm}
\vspace{0.5cm}
\caption{\label{fig:rhowall} Logarithm of density profiles $\rho_{wb}(z)$ of the big species $b$ of a binary hard-sphere mixture, with size ratio $q=0.5$, adsorbed at a planar hard wall. The bulk packing fraction $\eta_b$ is fixed at 0.1. The solid lines refer to DFT calculations and the dashed lines result from leading order asymptotics --- see text.
For a low packing fraction of the small species $\eta_s=0.1$ [curve (a)] the oscillations have a wavelength of about $\sigma_b$, the diameter of the big spheres whereas for a high packing fraction $\eta_s=0.15$ [curve (c)] the wavelength is about $\sigma_s$, the diameter of the small spheres. In the vicinity of the crossover, curve (b) with $\eta_s=0.125$, interference effects are clearly evident. In (a) and (c) a contribution from a single (dominant) pole, equation (\ref{eqn:decaywall2}), is used in the asymptotic expression whereas in curve (b) the contributions of the two poles are included, equation (\ref{eqn:decaywallcrossover}). The results in (b) and (c) are shifted vertically for clarity of display.}
\end{figure}

In our numerical calculations we consider the binary hard-sphere mixture for which we reported the crossover line in figure \ref{fig:crossoverline}, namely with a size ratio $q=0.5$. We determined the density profiles $\rho_{wi}(z)$ using Rosenfeld's fundamental measures DFT \cite{bib:rosenfeld89}, treating the planar wall as an external potential for the fluid mixture. Two types of wall-fluid potential were investigated. The first corresponds to a planar hard-wall:
\begin{equation}
V_{w}^i(z) =  \left\{
\begin{array}{cc}
\infty, \quad	& \quad z < R_i\\
0, \quad	& \quad \textrm{otherwise},
\end{array}
\right.
\label{eqn:potentialHW}
\end{equation}
for $i=s,b$. Here $z$ denotes the centre of a hard sphere, $R_b=\sigma_b/2$ and $R_s=\sigma_s/2$. In figure \ref{fig:rhowall} we show density profiles for the big spheres, $\rho_{wb}(z)$, for a fixed bulk (reservoir) packing fraction $\eta_b=0.1$ and three values of $\eta_s$, the bulk packing fraction of the smaller spheres. For $\eta_s=0.1$, curve (a), the oscillations have a wavelength of approximately $\sigma_b$ whereas for $\eta_s=0.15$, curve (c), the wavelength is approximately $\sigma_s = q \sigma_b$.  In curve (b), $\eta_s=0.125$, the density profile exhibits interference effects and two wavelengths are clearly evident. This value of $\eta_s$ lies close to the crossover value $\eta_s^{\ast} =0.126$ for this choice of $\eta_b$ --- see figure \ref{fig:crossoverline}. Structural crossover manifests itself in a very direct fashion. The other striking feature of these results is that the oscillatory behaviour predicted by leading order asymptotics sets in at {\em intermediate} distances, i.e.~for $z \gtrsim 2.5 \sigma_b$. This is illustrated by plotting alongside the DFT results the prediction from equation (\ref{eqn:decaywall2}), with $a_0$ and $a_1$ calculated from the zeros of $\hat D(k)$ in equation (\ref{eqn:denominator}) corresponding to the appropriate bulk state point $(\eta_b,\eta_s)$. As in subsection \ref{sec:theorybulk}, the input to $\hat D(k)$ are the pair direct correlation functions $\hat c_{ij}^{(2)}(k)$ obtained from Percus-Yevick theory. This is completely consistent with employing the Rosenfeld DFT since in bulk the latter {\em generates}, via functional differentiation, the Percus-Yevick $c_{ij}^{(2)}(r)$ \cite{bib:rosenfeld89,bib:grodon04}. 

The dashed lines in curves (a) and (c) of figure \ref{fig:rhowall} are results based on the single pole approximation, equation (\ref{eqn:decaywall2}), with the amplitude $B_b^w$ and phase $\Theta_b^w$ fitted to the DFT results at intermediate $z$. The agreement between the two sets of results for $\rho_{wb}(z)$ is excellent, apart from the region very close to the wall where many poles contribute. For curve (b), the state point close to crossover, we utilise a two pole approximation, equivalent to equation (\ref{eqn:decaycrossover}):
\begin{eqnarray}
\rho_{wi}(z) - \rho_i &\sim& B_i^w \exp{(-a_0 z)} \cos{(a_1 z - \Theta_i^w)} \nonumber \ \\
&&+ \tilde B_i^w \exp{(-\tilde a_0 z)} \cos{(\tilde a_1 z - \tilde \Theta_i^w)}, \quad z \to\infty
\label{eqn:decaywallcrossover}
\end{eqnarray}
$i=s,b$. Once again $a_0$, $\tilde a_0$, $a_1$ and $\tilde a_1$ are obtained from the zeros of $\hat D(k)$ at the given state point and now two amplitudes $B_b^w$ and $\tilde B_b^w$ and two phases $\Theta_b^w$ and $\tilde \Theta_b^w$ are fitted to the DFT data at intermediate $z$. The two-pole fit clearly provides an excellent description in the crossover regime. Overall the quality of the fits achieved by leading order asymptotics is no worse than is achieved by the corresponding fits to the test particle DFT results for $h_{ij}(r)$ obtained for size ratio $q=0.3$ --- see figure 6 of paper I. The implication of this observation is that the crossover features found in the DFT results should be clearly visible in computer simulations of hard wall-fluid density profiles which, owing to statistical considerations, are limited to small and intermediate $z$. 

The second type of planar wall that we consider is described by the wall-fluid potentials
\begin{equation}
\beta V_{\textsc{paw}}^i(z) =  \left\{
\begin{array}{lc}
\infty, \quad	& z < R_i\\
-\epsilon + \frac{2 \epsilon}{\lambda} (z-R_i) -\frac{\epsilon}{\lambda^2} (z-R_i)^2, \quad\quad& R_i < z < \lambda+R_i\\
0, \quad	& \textrm{otherwise},
\end{array}
\right.
\label{eqn:potentialPAW}
\end{equation}
where $\beta=(k_B T)^{-1}$ and $\epsilon>0$ is a dimensionless constant describing the depth of the attractive well, i.e.~$\beta V_{PAW}^i(R_i) = -\epsilon$. The range of the potential is given by the length scale $\lambda$ and the form is chosen so that the potential and its first derivative are zero at $z=\lambda + R_i$. In our calculations we set $\lambda = 1.25 \sigma_b$ and $\epsilon = 1$.

\begin{figure}
\centering\epsfig{file=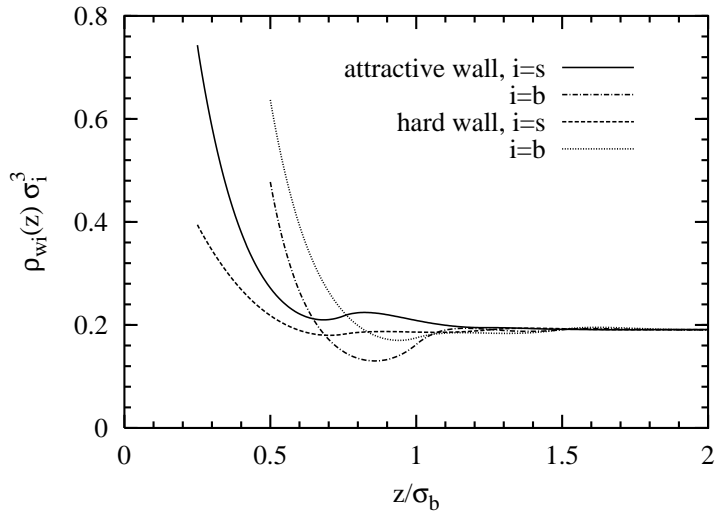,width=10cm}
\vspace{0.5cm}
\caption{\label{fig:rho_attr_lin} DFT results for the density profiles $\rho_{wi}(z)$ of the components $i=b,s$ of a hard-sphere mixture adsorbed at the planar attractive wall, described by equation (\ref{eqn:potentialPAW}), compared with the profiles at the planar hard wall, equation (\ref{eqn:potentialHW}), for the case $\eta_b=0.1$, $\eta_s=0.1$ and $q=0.5$. The additional short-ranged attraction of the wall ensures that the small particles ($s$) exhibit a higher density at contact, $\rho_{ws}(\sigma_s/2)$, and for $z$ close to the wall than they do at the hard wall. For the short distance scale shown here the density profiles clearly reflect the influence of the attraction.}
\end{figure}

\begin{figure}
\centering\epsfig{file=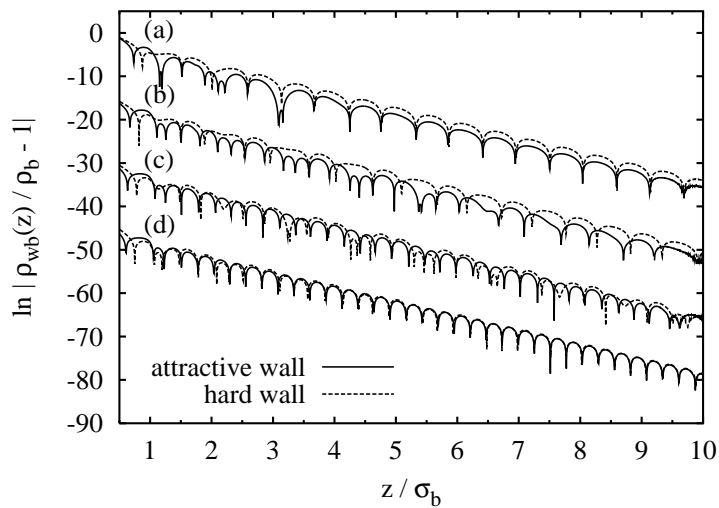,width=10cm}
\vspace{0.5cm}
\caption{\label{fig:rho_attr_log} As in figure \ref{fig:rho_attr_lin} ($q=0.5$, $\eta_b=0.1$) but now we plot the logarithm of the density profile $\rho_{wb}(z)$ of the big species $b$ over a much larger range of $z$. (a) $\eta_s=0.075$, (b) $\eta_s =0.1$, (c) $\eta_s=0.125$ (close to the crossover value) and (d) $\eta_s=0.15$. The full lines refer to results for the attractive wall, equation (\ref{eqn:potentialPAW}), and the dashed lines to those for the hard-wall, equation (\ref{eqn:potentialHW}). In (a) and (d) the pair of profiles are very similar for $z\gtrsim 3 \sigma_b$, whereas in (b) the asymptotic behaviour for the attractive wall does not appear to set in until $z\gtrsim 7 \sigma_b$. In (c) it is difficult to resolve two distinct wavelengths for the attractive wall. The curves are shifted vertically for clarity of display.}
\end{figure}

Including such an attractive component to the wall-fluid potentials will modify the density profiles in the vicinity of the wall but should not alter the fundamental character of the long ranged decay; leading order asymptotics should be just as accurate as in the case of the purely hard wall. Of course, the amplitude and phase in equation (\ref{eqn:decaywall2}) will be altered. In figure \ref{fig:rho_attr_lin} we show the density profiles $\rho_{wi}(z)$, $i=s,b$, of the binary hard-sphere mixture with $q=0.5$ calculated for the bulk state point $\eta_b=0.1$, $\eta_s=0.1$. Results for the hard-wall, described by equation (\ref{eqn:potentialHW}), are compared with those for the attractive wall, equation (\ref{eqn:potentialPAW}). It is energetically favourable to have a large number of particles in the attractive well. Since small spheres can pack better close to the wall than the larger ones it follows that the contact value, $\rho_s(\sigma_s/2)$, is considerably larger for small particles at the attractive wall than is the case for the same species at the hard-wall. Figure~\ref{fig:rho_attr_lin} demonstrates that adding an attractive piece to the hard-wall potential has a pronounced effect on the form of the density profiles of both species. From figure \ref{fig:rho_attr_lin} one cannot observe any commonality of structure between the profiles of big and small species or between profiles corresponding to the two choices of wall-fluid potentials. Rather it is necessary to study the density profiles at intermediate and long range in order to observe such common features. In figure \ref{fig:rho_attr_log} we plot the logarithm of the profile of the big spheres for $z$ up to $10\sigma_b$, the limit of our numerical accuracy. As was the case for the hard-wall, the asymptotic behaviour appears to set in at intermediate distances, $z\gtrsim 3 \sigma_b$. Away from crossover, e.g.~in curve (a) with $\eta_s=0.075$ and curve (d) with $\eta_s=0.15$ the results for the two types of wall-fluid potential are very close. For both types the wavelength of oscillations is close to $\sigma_b$ in (a) and close to $\sigma_s$ in (d). Clearly the asymptotic decay remains dictated by a single pole contribution when the wall exerts some finite ranged attraction on the fluid. For values of $\eta_s$ lying closer to crossover the situation is somewhat more complex. Curve (b) in figure \ref{fig:rho_attr_log}, for $\eta_s=0.1$, shows that the attractive wall potential influences strongly the form of the density profiles out to distances $z \sim 7 \sigma_b$ and the asymptotic behaviour, characterised by the wavelength $\sigma_b$, does not set in until larger $z$. For $\eta_s=0.125$, curve (c), there are clear differences between the results for the two different wall potentials extending out to $z\sim 10 \sigma_b$. It would appear that for the attractive wall the intermediate range behaviour of the profile is already dominated by the length scale $\sigma_s$, although crossover does not occur until the slightly larger value $\eta_s^{\ast}=0.126$. Clearly the amplitudes entering the two-pole approximation (\ref{eqn:decaywallcrossover}) depend sensitively on the choice of wall-fluid potential. Presumably for the attractive wall the amplitude $\tilde B_b^w$ corresponding to the pole with $\tilde a_1 = 2\pi/\sigma_s$ is much larger than $B_b^w$, that for the pole with $a_1 = 2\pi/\sigma_b$  whereas the imaginary part of the pole $a_0$ is slightly less than $\tilde a_0$.

We conclude that crossover in the wavelength of oscillations for density profiles is not restricted to purely repulsive walls but can also be observed when the wall exerts attractive forces on the fluid. 

Note that although we have focused on the profile of the big species, our results for the small species show equivalent features --- as is implied by equations (\ref{eqn:decaywall2}) and (\ref{eqn:decaywallcrossover}). We have also investigated density profiles for other size ratios $q$ and we find similar manifestations of crossover near the crossover point appropriate to the given value of $q$. Overall the behaviour is equivalent to that found for $h_{bb}(r)$ in paper I.

%%%%%%%%%%%%%%%%%%%%%%%%%%%%%%%%%%%%%%%%%%%%%%%%%%%%%%%%%%%%%%%%%%%%%%%%%%%%%%%%
%%%%%%%%%%%%%%%%%%%%%%%%%%%%%%%%%%%%%%%%%%%%%%%%%%%%%%%%%%%%%%%%%%%%%%%%%%%%%%%%
%%%%%%%%%%%%%%%%%%%%%%%%%%%%%%%%%%%%%%%%%%%%%%%%%%%%%%%%%%%%%%%%%%%%%%%%%%%%%%%%
\section{Depletion Potentials and Solvation Forces}

In this section we turn attention to the asymptotic decay of two, closely related, properties of liquids. These are i) the depletion or solvent mediated potential, $W(r)$, which arises between two big particles immersed in a solvent and ii) the solvation force, $f_s(L)$, which is the excess pressure arising from confining the solvent between two planar walls. We present evidence for structural crossover in $W(r)$, at large particle separations $r$, and in $f_s(L)$, at large wall separations $L$, when the solvent is the binary hard-sphere mixture described in earlier sections.

%%%%%%%%%%%%%%%%%%%%%%%%%%%%%%%%%%%%%%%%%%%%%%%%%%%%%%%%%%%%%%%%%%%%%%%%%%%%%%%%
\subsection{Depletion or Solvent Mediated Potential}
\label{sec:depletion}

Consider two big particles of species 3 immersed in a solvent. Upon integrating out the degrees of freedom of the solvent atoms or molecules one obtains an effective two-body potential
\begin{equation}
\phi_{33}^{\mathit{eff}}(r) \equiv \phi_{33}(r) + W(r),
\end{equation}
where $\phi_{33}(r)$ is the bare interaction between the solute particles and $W(r)$ is the solvent mediated potential arising from solute-solvent and solvent-solvent interactions. $W(r)$ depends on the thermodynamic state of the solvent. A well-known example is the case of two hard-spheres, modelling colloids, immersed in a sea of ideal, non-interacting particles, modelling 'ideal polymer'. The exclusion of the 'polymer' from the hard spheres gives rise to a depletion zone between the spheres and the resulting $W(r)$ is the celebrated depletion potential introduced by Asakura and Oosawa \cite{bib:asakura54,bib:asakura58} and by Vrij \cite{bib:vrij76}. For the ideal solvent $W(r)$ is attractive and monotonically increasing with $r$ up to a certain finite separation beyond which it vanishes. When the solvent is a hard-sphere fluid and the solute particles are again (big) hard-spheres, $W(r)$ exhibits depletion driven attraction at small separations and oscillatory decay at intermediate and long range. The oscillations arise from the packing of the small spheres that constitute the solvent and their wavelength is determined by the diameter of the small species, e.g.~Ref.~\cite{bib:roth00}. 
   
Here we investigate the depletion potential when the solvent is the binary mixture of $s$ and $b$ hard-spheres. On intuitive grounds one might expect the oscillatory structure in $W(r)$ to reflect the form of correlations in the bulk binary mixture and, therefore, to show crossover behaviour. In order to make this explicit we generalise the argument given in Ref.~\cite{bib:roth00} for a one-component solvent to a two-component solvent. For a ternary mixture of species $s,b$ and $3$ the Fourier transform of the total correlation function $\hat h_{33}(k)$ is given by an expression equivalent to equation (\ref{eqn:FourierOZ}). In the limit where the density of species $3$ vanishes, $\rho_3 \to 0$, the denominator should reduce to that in equation (\ref{eqn:denominator}), as was the case for wall-particle correlations in equation (\ref{eqn:FourierOZwall}), so the asymptotic decay of $h_{33}(r)$ will be determined by the zeros of $\hat D(k)$ at the appropriate state point. In the same limit we have 
\begin{equation}
h_{33}(r) + 1 = g_{33}(r) = \exp{[-\beta \phi_{33}^{\mathit{eff}}(r)]}.
\label{eqn:h33}
\end{equation}
It follows that for large $r$, where we can linearise the exponential, 
\begin{equation}
-\beta W(r) \sim h_{33}(r) \sim \frac{A_{33}}{r} \exp{(-a_0 r)} \cos{(a_1 r -\Theta_{33})}, \quad r\to \infty. 
\label{eqn:deplpotdecay}
\end{equation}
Note that we assume the bare potential $\phi_{33}(r)$ is of shorter range than $W(r)$. The same LOP that determines the decay of $h_{ij}(r)$, $i=s,b$, in the bulk mixture must determine the decay of the depletion potential. Of course, the amplitude $A_{33}$ and phase $\Theta_{33}$ depend upon the solute-solvent interactions.

\begin{figure}
\centering\epsfig{file=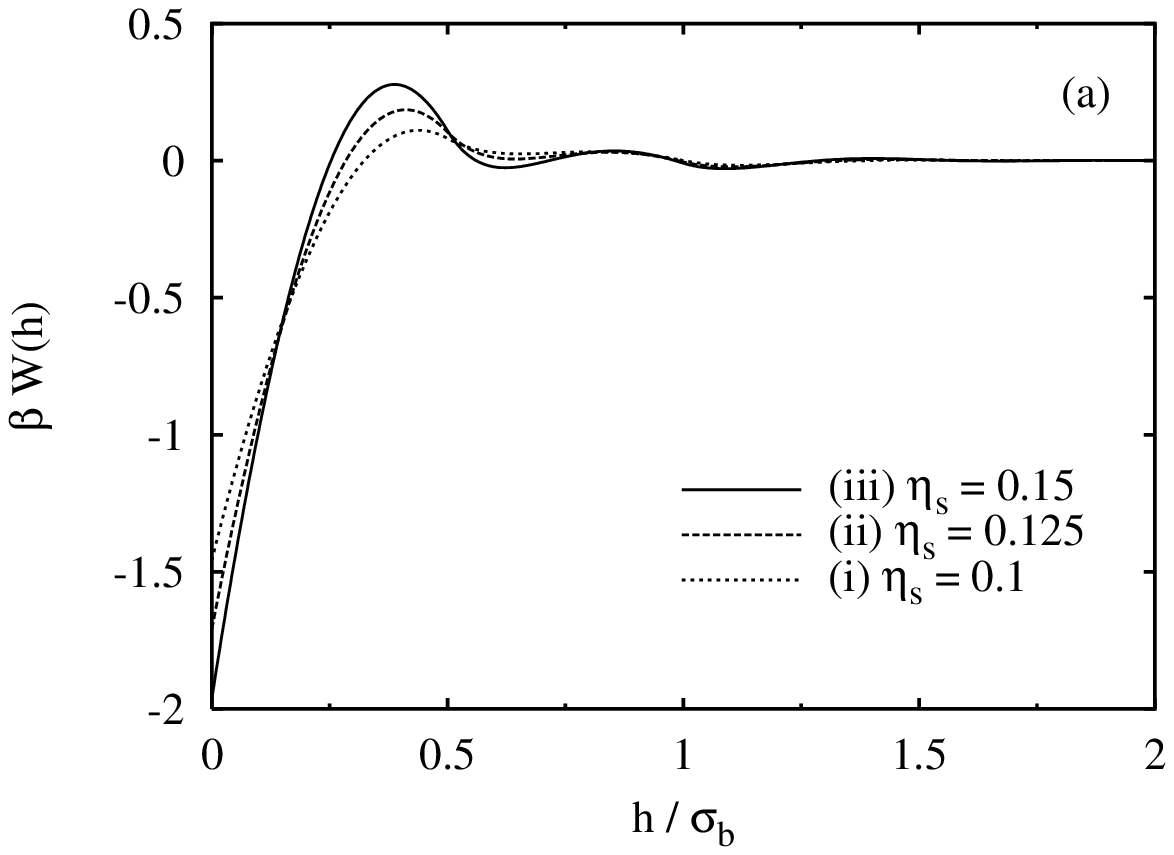,width=10cm}
\centering\epsfig{file=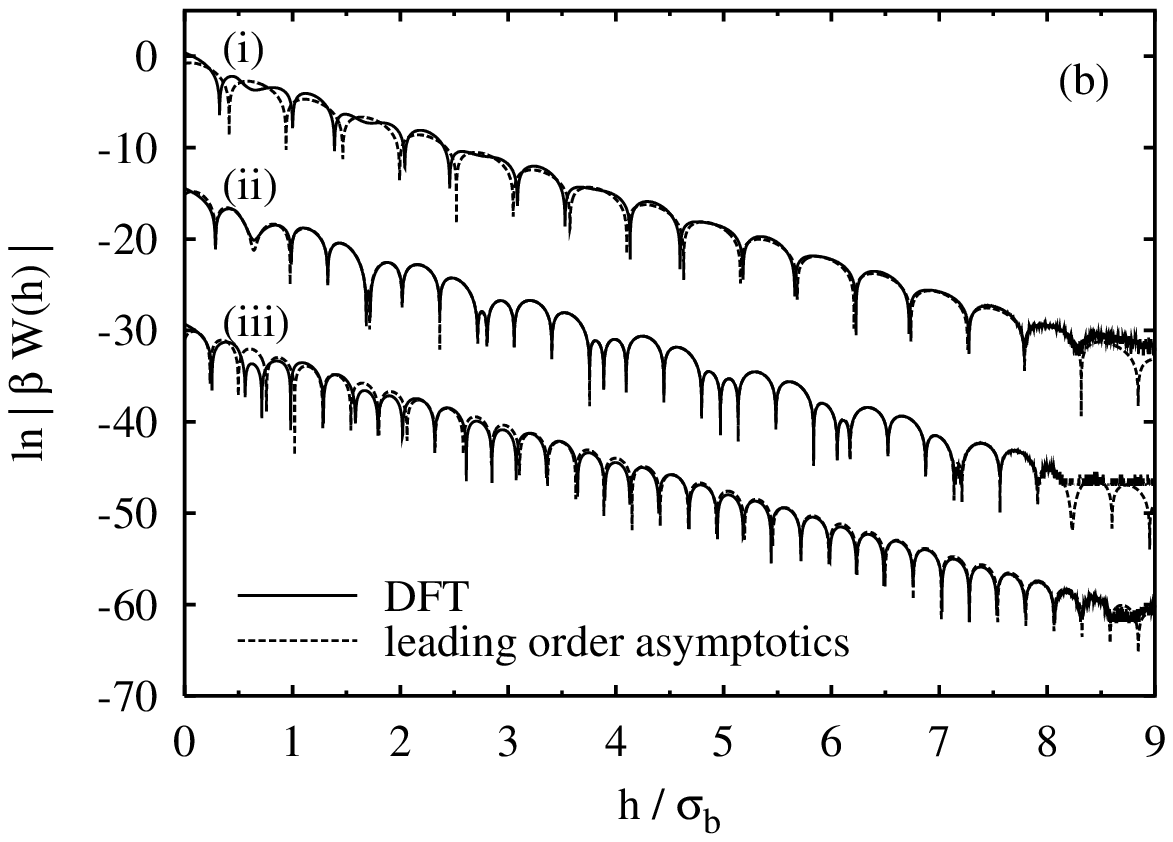,width=10cm}
\vspace{0.5cm}
\caption{\label{fig:depletion} Depletion potential $W(h)$ between two hard-spheres of diameter $\sigma_3=5\sigma_s$ immersed in a binary mixture of hard spheres with diameters $\sigma_s$ and $\sigma_b=2\sigma_s$, i.e.~$q=0.5$. $h = r - \sigma_3$ is the separation between the surfaces of the two big spheres of species 3 at distance $r$. The packing fraction of species $b$ is $\eta_b=0.1$. DFT results are shown in (a) for three different values of $\eta_s$. Increasing $\eta_s$ reduces the width of the attractive well and increases its depth at contact $h=0$. In (b) we plot the logarithm of $|\beta W(h)|$ over a larger range of separations. For $\eta_s=0.1$ [curve (i)] the oscillations have a wavelength of about $\sigma_b$ whereas for $\eta_s=0.15$ [curve (iii)] the wavelength is about $\sigma_s$. In curve (ii) $\eta_s=0.125$, very close to crossover, interference effects are evident. The solid lines in (b) are the DFT results and the dashed lines result from leading order asymptotics --- see text. The results (ii) and (iii) are shifted vertically for clarity of display. These results demonstrate the same crossover behaviour as the density profiles in figure \ref{fig:rhowall}.}
\end{figure}

In order to determine whether structural crossover can be observed in the depletion potential we performed calculations of $W(r)$ using the particle insertion method employed previously in DFT studies for one-component hard-sphere solvents \cite{bib:roth00}. We first fix a single sphere of diameter $\sigma_3$ at the origin and make this an external potential for the fluid mixture of $s$ and $b$ hard-spheres. As discussed earlier, the density profiles of both $s$ and $b$ particles, obtained using the Rosenfeld DFT, display structural crossover as $r \to \infty$. In a second step, a second particle of species $3$ is inserted into the inhomogeneous mixture of $s$ and $b$ and input from the three component DFT is used to calculate $W(r)$ \cite{bib:roth00}. The results shown in figure \ref{fig:depletion} refer to the hard-sphere mixture with $q=0.5$ and $\eta_b=0.1$ considered earlier. We set $\sigma_3=5 \sigma_s$. At short distances from contact ($h=0$ in figure \ref{fig:depletion}, where $h=r-\sigma_3$ is the separation between the surfaces of the two big spheres of species $3$) the depletion potential is attractive whereas oscillatory structure develops for larger distances. Results are presented for the same three choices, $\eta_s=0.1$, $\eta_s=0.125$ and $\eta_s=0.15$, employed in figure \ref{fig:rhowall}. From figure \ref{fig:depletion}(a) we note that the width of the attractive depletion well seems to be determined by the size of the majority species in the mixture. Thus, for $\eta_s=0.1$ the big spheres dominate and the width of the attractive well is about $0.4 \sigma_b$. On the other hand for $\eta_s=0.15$, where the smaller species dominates, the attractive well is deeper and narrower and the depletion potential exhibits a pronounced potential maximum near $h=0.4\sigma_b$.

In figure \ref{fig:depletion}(b) we show the same data plotted as $\ln{|\beta W(h)|}$ in order to expose the nature of the oscillatory decay. For $\eta_s=0.1$ the wavelength is about $\sigma_b$ whereas for $\eta_s=0.15$ this is about $\sigma_s=\sigma_b / 2$. Interference is clearly visible for $\eta_s=0.125$, very close to the crossover value. The sequence of curves is very similar to that displayed in figure \ref{fig:rhowall} for the density profiles at a hard planar wall. We confirmed that the crossover mechanism is identical by plotting (as dashed lines) alongside the DFT results those obtained from the single pole approximation (\ref{eqn:deplpotdecay}) in which $a_0$ and $a_1$ refer to the LOP and are equal to the values employed in figure \ref{fig:rhowall}. For $\eta_s=0.1$ and $0.15$, i.e.~curves (i) and (iii), the single pole approximation gives an excellent account of the DFT results for $h \gtrsim 3.5 \sigma_b$. Close to the crossover point, curve (ii) with $\eta_s=0.125$, two LOPs are required, i.e.~we employ the analogue of equation (\ref{eqn:decaycrossover}). As in figure \ref{fig:rhowall}, we fit amplitudes and phases at intermediate separations. By comparing figure \ref{fig:depletion}(b) with figure \ref{fig:rhowall} we see that crossover is just as evident in the depletion potential as in the density profiles at a planar wall. Note that numerical noise sets into our results for $h\gtrsim 8 \sigma_b$.

%%%%%%%%%%%%%%%%%%%%%%%%%%%%%%%%%%%%%%%%%%%%%%%%%%%%%%%%%%%%%%%%%%%%%%%%%%%%%%%%
\subsection{Solvation Force}
\label{sec:solvationforce}

The solvation force is defined \cite{bib:evans90,bib:evans87} for a planar slit of width $L$ as
\begin{equation}
f_s(L) \equiv - \frac{1}{A}\left( \frac{\partial \Omega}{\partial L}\right)_{\{\mu_i\},T,A} - \ p,
\label{eqn:solvationforce}
\end{equation}
where $\Omega$ is the grand potential of the fluid mixture and $p$ is the pressure of a homogeneous mixture with the (reservoir) chemical potentials $\{\mu_i\}$, $i=s,b$, and temperature $T$. We consider the limit where the area of the walls \mbox{$A\to \infty$}. $f_s(L)$ is then the excess pressure resulting from confining the mixture in the slit; in the limit $L\to\infty$, $f_s(L)\to 0$. Here we consider only fluid-fluid and wall-fluid contributions to the solvation force, not the direct interaction between the the walls \cite{bib:evans87}. In the asymptotic regime, $L\to\infty$, one might expect the decay of $f_s(L)$ to be determined by the properties of the bulk mixture described by the given $\{\mu_i\}$, $T$. However, it is not immediately obvious that the thermodynamic quantity $f_s(L)$ should decay in the same fashion as the bulk pair correlation functions $h_{ij}(r)$, i.e.~with the same exponential decay length $a_0^{-1}$ and wavelength $2\pi/a_1$.

There are several ways of establishing this result. The most intuitive argument is to consider the ternary mixture described in the previous subsection, with $\rho_3\to 0$ and, in addition, to allow the diameter of species 3 to become macroscopically large. The effective potential between the macroscopic particles will still decay as in equation (\ref{eqn:deplpotdecay}). Attard and co-workers \cite{bib:attard91} placed this argument in the framework of the wall-particle OZ equations (see subsection \ref{sec:theorywall}) and showed that the decay of the excess interaction free energy per unit area should be governed by the zeros of $\hat D(k)$, provided that the interatomic forces are sufficiently short-ranged. Such a result is also implicit in earlier integral equation studies by D.~Henderson and co-workers, e.g.~Ref.~\cite{bib:dhenderson86}.

In the particular case of perfectly hard-walls, for which equation (\ref{eqn:solvationforce}) reduces to the exact expression \cite{bib:henderson86}
\begin{equation}
\beta f_s(L) = \sum_{i=b,s} [\rho_{i,L}(0^+) - \rho_{i,\infty}(0^+)],
\label{eqn:contacttheorem}
\end{equation}
where $\rho_{i,L}(0^+)$ denotes the one-body density of species $i$ at contact with one wall of the slit of width $L$ and $\rho_{i,\infty}(0^+)$ is the same quantity for infinite wall separations, there are additional arguments \cite{bib:evans93} to support the contention that
\begin{equation}
f_s(L) \sim F \exp{(-a_0 L )} \cos{(a_1 L - \Theta_F)}, \quad L\to \infty
\label{eqn:sfdecay}
\end{equation}
where $F$ and $\Theta_F$ are the amplitude and phase, respectively. Both quantities depend on the wall-fluid potentials.

In the following we study the binary-hard sphere mixture confined by two planar hard walls described by the potential
\begin{equation}
V_{slit}^{i}(z) =  \left\{
\begin{array}{cc}
\infty, \quad	& \quad z \leq R_i \ \textrm{and} \ L-R_i \leq z\\
0, \quad	& \quad R_i < z < L-R_i,
\end{array}
\right.
\label{eqn:potslit}
\end{equation}
which is the generalisation of equation (\ref{eqn:potentialHW}). In our calculations we use the external potential equation (\ref{eqn:potslit}) as input into the Rosenfeld fundamental measures DFT. We obtain the contact values $\rho_{i,L}(0^{+})$ from the density profiles $\rho_{i,L}(z)$ resulting from minimising the grand potential functional $\Omega[\{\rho_i(z)\}]$. Note that the profiles obtained from the DFT satisfy the contact value theorem, i.e.~the solvation force obtained from equation (\ref{eqn:contacttheorem}) is consistent with that obtained from differentiating w.r.t.~$L$ the equilibrium grand potential $\Omega$ resulting from minimising the functional. Considerable computational effort is required to produce sufficient values of the solvation force that we can have smooth curves as a function of the separation $L$. In order to investigate crossover behaviour we focused on the same mixture with $q=0.5$ and reservoir packing fraction $\eta_b=0.1$ and varied $\eta_s$ over a similar range to that considered in earlier sections. 

\begin{figure}
\centering\epsfig{file=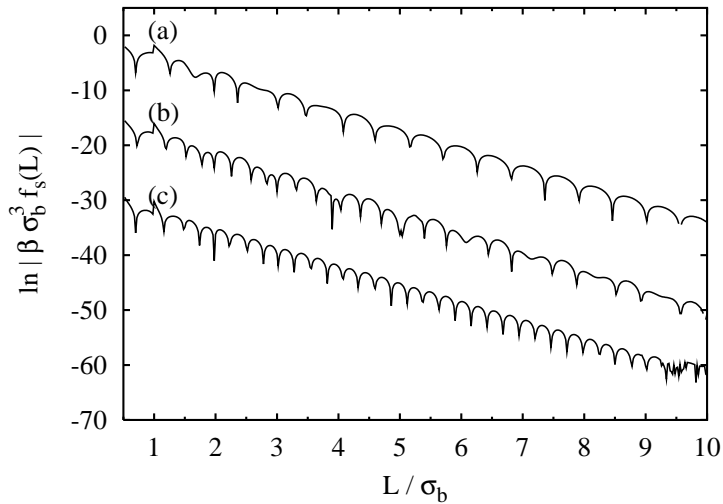,width=10cm}
\vspace{0.5cm}
\caption{\label{fig:solvationforce_log} Logarithm of the solvation force $f_s(L)$ calculated from DFT for a binary hard-sphere mixture, with size ratio $q=0.5$ and $\eta_b=0.1$, confined between two planar hard walls. Upon increasing the packing fraction of the small particles from $\eta_s=0.07$ [curve (a)] to $\eta_s=0.15$ [curve (c)] we find a crossover in the wavelength of oscillations. Interference effects can be observed at intermediate separations $L$ for $\eta_s=0.11$ [curve (b)]. Curves (b) and (c) have been shifted vertically for clarity of display.}
\end{figure}

In figure \ref{fig:solvationforce_log} we plot the logarithm of $|\beta \sigma_b^3 f_s(L)|$ for three values of $\eta_s$. For a small packing fraction, curve (a) $\eta_s=0.07$, the oscillations in $f_s(L)$ have a wavelength of about $\sigma_b$ for $L \gtrsim 3.5\sigma_b$. At a high packing fraction, curve (c) $\eta_s=0.15$, the wavelength of oscillations is given by the diameter of the small spheres $\sigma_s$ for $L\gtrsim 2 \sigma_b$. At intermediate values of $\eta_s$ we observe interference effects in the oscillatory decay. In curve (b), where $\eta_s=0.11$, the oscillations at small $L$ have a wavelength of about $\sigma_s$ whereas at intermediate $L$ interference effects resulting from two distinct contributions are apparent. Finally for large $L$, the oscillations have a wavelength of about $\sigma_b$. Clearly confinement results in the small spheres imposing their length scale on the oscillations in $f_s(L)$ at small $L$ and restricts interference effects to larger $L$. Recall that the crossover value for this choice of $\eta_b$ is $\eta_s^{\ast}=0.126$. Comparing the results in figure \ref{fig:solvationforce_log} with the density profiles $\rho_{wb}(z)$ for a single planar hard-wall in figure \ref{fig:rhowall} we find close similarities. In particular, curves (b) and (c) in figure \ref{fig:solvationforce_log} provide compelling evidence that crossover has occurred in the form of oscillatory decay of $f_s(L)$ for packing fractions between $\eta_s=0.11$ and $\eta_s=0.15$. Note that there is some noise in the latter results for $L \gtrsim 9 \sigma_b$, the limit of our numerical accuracy.

It is important to recognise that there is no jump in the solvation force, or its derivatives at the crossover line. Near crossover two LOPs contribute to $f_s(L)$, in the manner of equations (\ref{eqn:decaycrossover}) and (\ref{eqn:decaywallcrossover}), and the real and imaginary parts of both poles vary smoothly with state point.

%%%%%%%%%%%%%%%%%%%%%%%%%%%%%%%%%%%%%%%%%%%%%%%%%%%%%%%%%%%%%%%%%%%%%%%%%%%%%%%%
%%%%%%%%%%%%%%%%%%%%%%%%%%%%%%%%%%%%%%%%%%%%%%%%%%%%%%%%%%%%%%%%%%%%%%%%%%%%%%%%
%%%%%%%%%%%%%%%%%%%%%%%%%%%%%%%%%%%%%%%%%%%%%%%%%%%%%%%%%%%%%%%%%%%%%%%%%%%%%%%%
\section{Polydisperse Hard-Sphere Mixtures}
\label{sec:polydispersity}

From an experimental point of view it is important to keep in mind that systems of colloidal particles with identical size (or shape) cannot be produced. It is not possible to produce a perfect binary mixture; there will always be a distribution of size ratios in even the best bimodal situation. Here we examine the influence of polydispersity on the structural crossover which was clearly manifest for a mixture of hard-spheres with a unique size ratio. The Rosenfeld DFT can treat mixtures with an arbitrary number of different components (species). However, we restrict consideration to a finite number of different particle sizes and replace each of the particle types big, $b$, and small, $s$, by a 21-component hard-sphere mixture.

\begin{figure}
\centering\epsfig{file=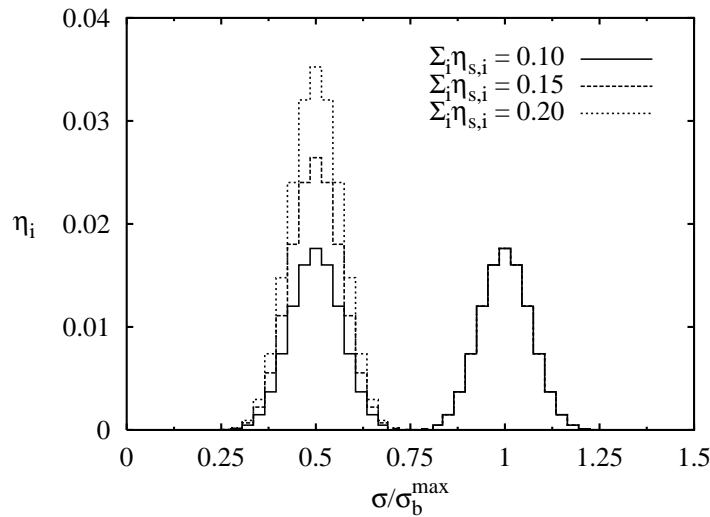,width=10cm}
\vspace{0.5cm}
\caption{\label{fig:distribution} Distributions of the components in a bulk polydisperse hard-sphere mixture plotted as the ratio of the hard-sphere diameter $\sigma$ to $\sigma_b^{max}$, the diameter of the component with maximum packing fraction in the distribution of the bigger spheres. 
Each binomial distribution consists of 21 components and the sum of packing fractions of the bigger components $\sum_i \eta_{b,i}=0.1$. The sum of the smaller components $\sum_i \eta_{s,i}$ takes values 0.1, 0.15, or 0.2. Note that the ratio $\sigma_s^{max}/\sigma_b^{max} = 0.5$; where $\sigma_s^{max}$ is the diameter of the component with maximum packing fraction in the distribution of the smaller spheres.}
\end{figure}

\begin{figure}
\centering\epsfig{file=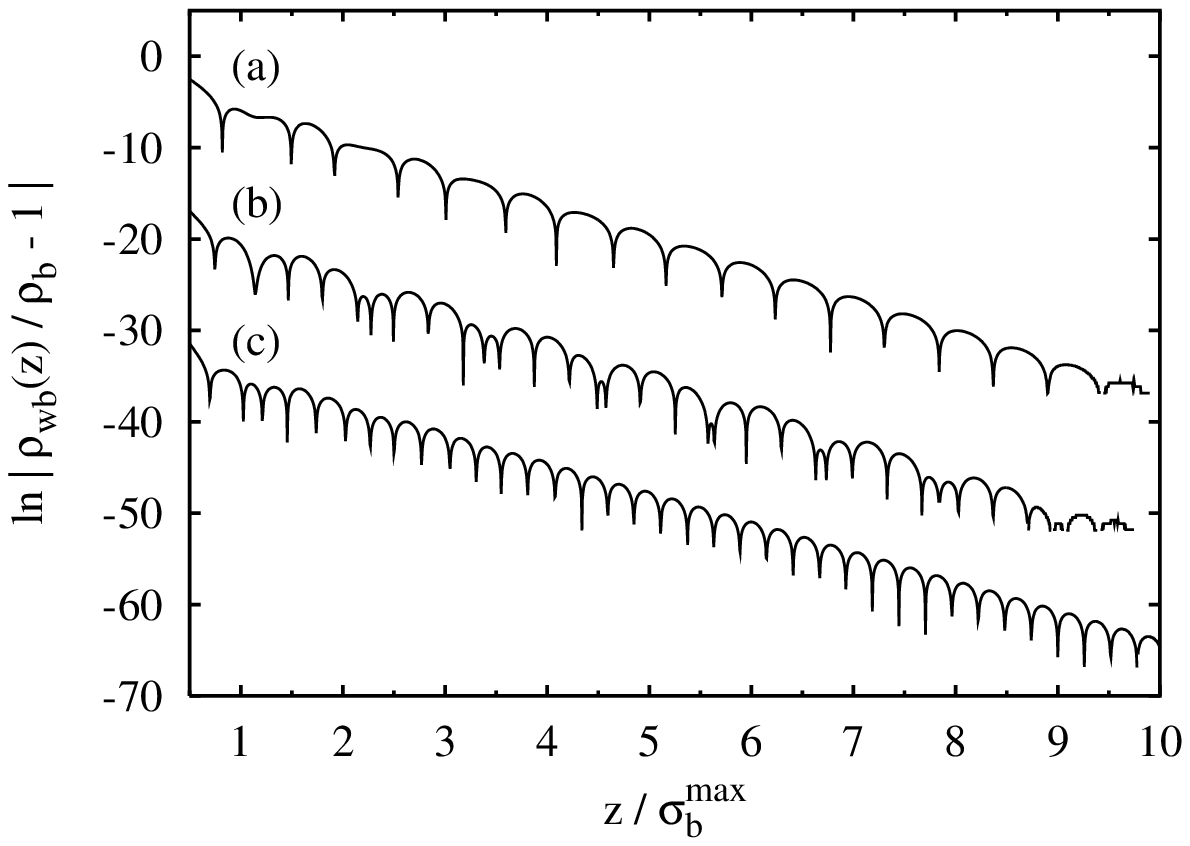,width=10cm}
\vspace{0.5cm}
\caption{\label{fig:rhowallpoly} Logarithm of density profiles of component $b$ in a polydisperse hard-sphere mixture,  with the particle size distributions shown in figure \ref{fig:distribution}, adsorbed at a planar hard-wall. The component $b$ considered here refers to that with $\sigma=\sigma_b^{max}$, the diameter corresponding to the maximum packing fraction in the distribution of the bigger spheres. For all three curves $\sum_i \eta_{b,i}=0.1$. In curve (a) the sum of packing fractions of the smaller components $\sum_i \eta_{s,i}=0.1$ and the wavelength of oscillations is about $\sigma_b^{max}$. By contrast, in curve (c), where $\sum_i \eta_{s,i}=0.2$, the wavelength is about $\sigma_b^{max}/2$. In the intermediate case, curve (b), with $\sum_i \eta_{s,i}=0.15$ we find interference effects. Results should be compared with these in figure \ref{fig:rhowall} for a binary mixture with $q=0.5$.}
\end{figure}

The size distributions are chosen to be binomial and are shown in figure \ref{fig:distribution}. In the case we consider the ratio of diameters $\sigma_s^{max}/\sigma_b^{max}$, corresponding to the positions of the two maxima, is 0.5. The sum of the packing fractions, $\sum_i \eta_{b,i}$ of the big particles is fixed at 0.1 while that of the smaller particles takes values $\sum_i \eta_{s,i}=0.1$, $0.15$, and $0.2$ in order to compare with the results from earlier sections. We treat rather broad distributions in order to capture effects which might arise in experimental situations. The width of both distributions is 0.6 and there is a very small overlap --- see figure \ref{fig:distribution}. 
Figure \ref{fig:rhowallpoly} shows the density profiles $\rho_{wb}(z)$ of the species with diameter $\sigma_b^{max}$ for polydisperse hard-sphere mixtures adsorbed at a planar hard wall.
The Rosenfeld grand-potential functional is minimised and the resulting Euler-Lagrange equations for the profiles of the multi-component mixture are solved imposing the boundary conditions that the reservoir packing fractions are those given in figure \ref{fig:distribution}. The results for $\rho_{wb}(z)$ are very similar to those in figure \ref{fig:rhowall} for the binary hard-sphere mixture with size ratio $q=0.5$. In curve (a), $\sum_{i}\eta_{s,i}=0.1$, the wavelength of the oscillatory decay is characterised by the big spheres whose reservoir packing fraction $\eta_{b,i}$ is a maximum. The small particles determine the wavelength of the oscillations in curve (c), $\sum_{i}\eta_{s,i}=0.2$. Curve (b),  $\sum_{i}\eta_{s,i}=0.15$, clearly displays interference effects. If we take the presence of the latter to signal crossover then this is occurring for a larger (total) small sphere packing fraction than is the case for the binary mixture, where crossover occurs at $\eta_s^{\ast} = 0.126$. This observation is consistent with results from paper I where we found that for $\eta_b=0.1$ the crossover value of $\eta_s$ increases as $q$ increases or decreases (see figure 5 of I), i.e.~the minimum value of the crossover packing fraction occurs near $q=0.5$. In the polydisperse mixture one is sampling component pairs with size ratios different from $q=0.5$. 

It is striking that for such broad distributions there is still a clear signature of crossover in the density profile and we shall return to this point in the discussion. 

%%%%%%%%%%%%%%%%%%%%%%%%%%%%%%%%%%%%%%%%%%%%%%%%%%%%%%%%%%%%%%%%%%%%%%%%%%%%%%%%
%%%%%%%%%%%%%%%%%%%%%%%%%%%%%%%%%%%%%%%%%%%%%%%%%%%%%%%%%%%%%%%%%%%%%%%%%%%%%%%%
%%%%%%%%%%%%%%%%%%%%%%%%%%%%%%%%%%%%%%%%%%%%%%%%%%%%%%%%%%%%%%%%%%%%%%%%%%%%%%%%
\section{Structural Crossover in One and Two-Dimensional Homogeneous Fluids}
\label{sec:rods}

All the results that we have presented so far refer to three-dimensional hard-sphere mixtures. Moreover these have been obtained by approximate methods, i.e.~via DFT or integral-equation approaches. In this section we enquire whether structural crossover manifests itself in lower dimensional fluid mixtures. There are two distinct reasons for considering such systems. In one-dimension the pair direct correlation functions $c_{ij}^{(2)}(r)$ can be calculated exactly for a homogeneous mixture of hard-rods of different lengths which implies that the poles of $\hat h_{ij}(k)$ can be determined exactly. The reason why two-dimensional systems are of interest is that real-space experimental techniques for colloidal fluids often restrict the determination of the pair correlation functions to two dimensions \cite{bib:brunner02,bib:klein02,bib:brunner03}.

\subsection{One-Dimensional Mixtures of Hard-Rods}
\label{sec:onedim}

One-dimensional hard-rod fluids occupy a special position in the theory of liquids since many of their equilibrium properties can be calculated exactly for both homogeneous and inhomogeneous cases. In particular the exact intrinsic free energy functional is known for both the one component fluid \cite{bib:percus82} and mixtures of hard-rods \cite{bib:vanderlick89}. For our present purposes it is sufficient to note that Rosenfeld's fundamental measure DFT reproduces the exact functional when applied to the one-dimensional inhomogeneous hard-rod system \cite{bib:rosenfeld89,bib:rosenfeld90}. The pair direct correlation functions are given by an expression equivalent to equation (11) in I but with different weight functions and a different excess free energy density $\beta \Phi$. Their (one dimensional) Fourier transforms $\tilde c_{ij}^{(2)}(k)$ are given in Ref.~\cite{bib:rosenfeld90}.

\begin{figure}
\centering\epsfig{file=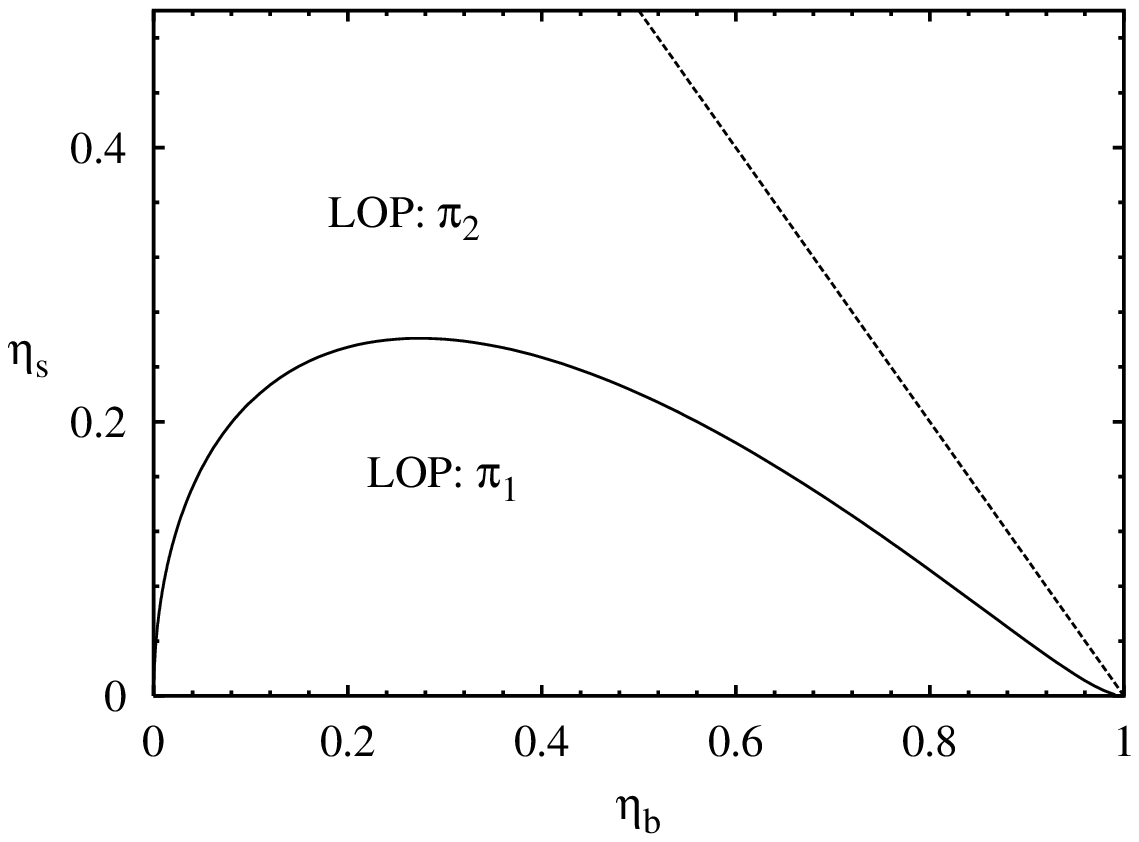,width=10cm}
\vspace{0.5cm}
\caption{\label{fig:crossoverline1d} Crossover line for a one-dimensional binary hard-rod mixture with size ratio $q=0.5$. On each side of the line the asymptotic behaviour is dominated by different leading-order poles $\pi_1$ and $\pi_2$. When $\pi_1$ dominates the wavelength of oscillations in $h_{ij}(r)$ is $\sim \sigma_b$ whereas when $\pi_2$ dominates it is $\sim 0.5\sigma_b$. These results were obtained using the exact result for $\tilde c_{ij}^{(2)}(r)$ to determine the poles. The dashed line corresponds to total packing fraction $\eta = \eta_b+\eta_s =1$.}
\end{figure}

We used these results to obtain the poles of $\tilde h_{ij}(k)$. These are given by the zeros of an expression equivalent to that for $\hat D(k)$ in equation (\ref{eqn:denominator}) since the OZ equations can still be written in the form of equation (\ref{eqn:FourierOZ}). The pattern of poles is similar to that obtained for hard-sphere mixtures in the Percus-Yevick approximation \cite{bib:grodon04}. In figure \ref{fig:crossoverline1d} we plot the crossover line calculated for size ratio $q=0.5$. The packing fractions are given by $\eta_i = \rho_i \sigma_i$, $i=s,b$, in one dimension; $\sigma_i$ is the length of the rod of species $i$. Crossover occurs from the pole $\pi_1$ to the pole $\pi_2$ in a similar fashion to that described in figures \ref{fig:crossoverline} and \ref{fig:fundamentalplot} for hard-spheres. Moreover the shape of the crossover line is reminiscent of that in figure \ref{fig:crossoverline}. Of course the line can extend to $\eta_b = 1$ since there is no freezing transition in one-dimension and the maximum packing fraction is unity.

These results demonstrate that crossover occurs in an exactly solvable model. In the vicinity of a crossover point we find similar features in the total pair correlation functions as those found in the three dimensional hard-sphere mixtures. Recall that for hard-rod mixtures it is possible to calculate wall-fluid density profiles and the solvation force exactly (to arbitrary numerical precision) which allows one to ascertain precisely the manifestation of structural crossover in these quantities \cite{bib:comment1d}.

\subsection{Two-Dimensional Mixtures of Hard-Disks}

Theories of liquids are, generally speaking, more difficult to construct and implement in two dimensions than in one or three dimensions. This is certainly the case for the fundamental measure DFT that we have used throughout this paper. Whereas in $d=1$ the exact functional for a hard-rod mixture \cite{bib:vanderlick89,bib:rosenfeld90} is known and in $d=3$ very accurate versions of FMT for hard-sphere mixtures \cite{bib:rosenfeld89,bib:roth02,bib:yu02} are available, some pre-requisites for FMT-like functionals are not fulfilled in $d=2$ \cite{bib:rosenfeld90}. The starting point for FMT is the deconvolution of the Mayer-$f$ function \cite{bib:rosenfeld89}, that describes the interaction between two particles, into geometrical weight functions. In even dimensions, however, this deconvolution can be achieved only with an infinite number of weight functions. Hence, in practice, the deconvolution of the Mayer-$f$ function in $d=2$ can be carried out only approximately \cite{bib:rosenfeld90}. For this reason the FMT results can be less accurate than in $d=3$, although there have been few careful tests of the theory for mixtures. 

On the other hand, the study of correlation functions in $d=2$ is very interesting from an experimental point of view. Recent improvements in techniques of video microscopy enable accurate measurement of pair correlation functions of colloidal suspensions to be performed in real space \cite{bib:brunner02,bib:klein02,bib:brunner03}. In these experiments, the colloids are restricted to move in a plane by the use of light fields, i.e.~the system is effectively two dimensional. Experiments with one-component colloidal suspensions have demonstrated that the pair correlation function can be measured very accurately at small and intermediate separations and, in some cases, even at fairly long range. An extension of such experiments to binary colloidal mixtures should be straightforward.

\begin{figure}
\centering\epsfig{file=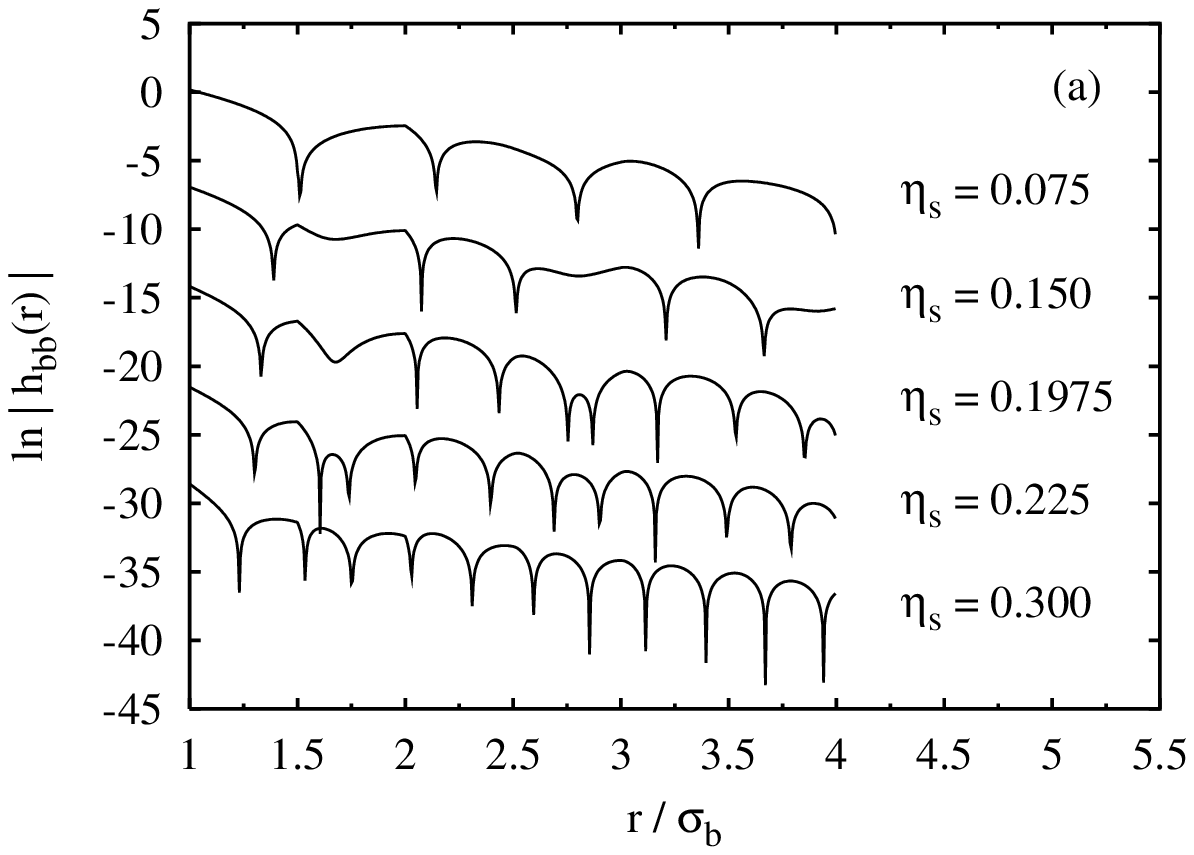,width=10cm}
\centering\epsfig{file=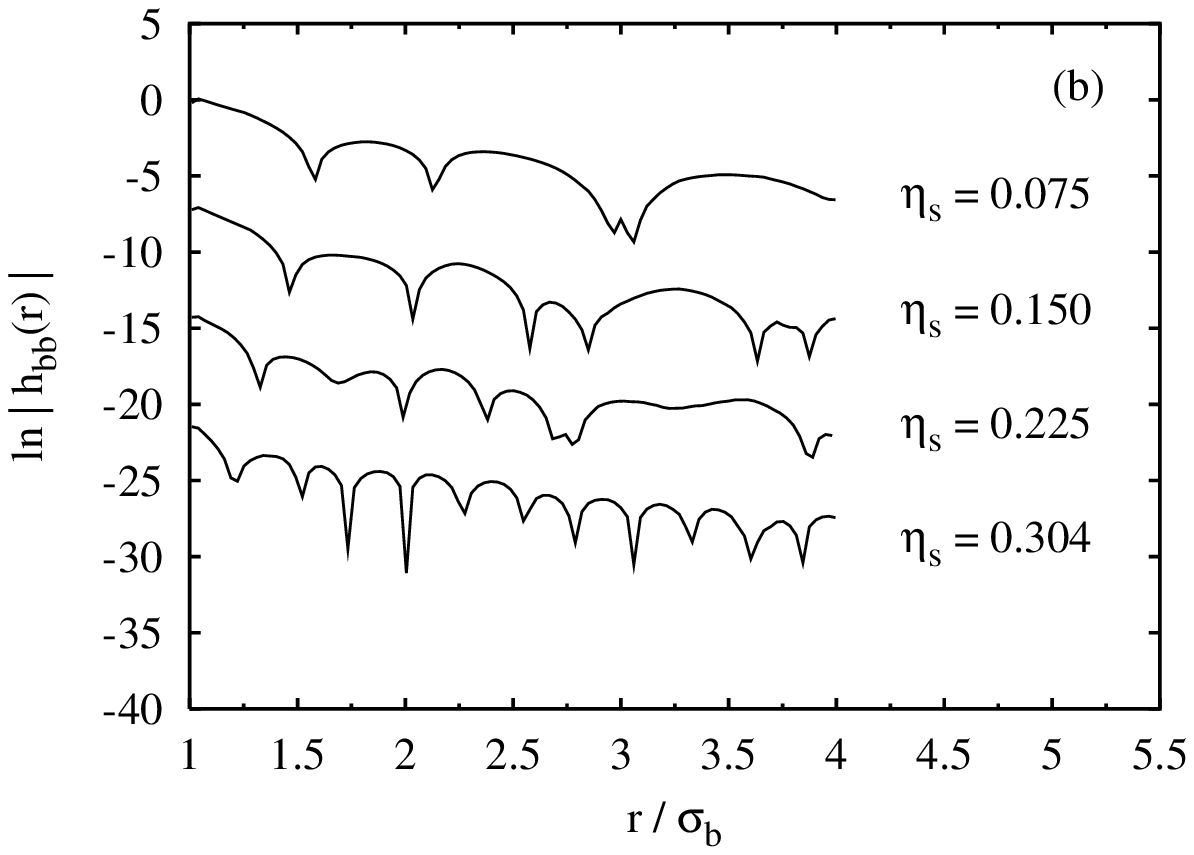,width=10cm}
\vspace{0.5cm}
\caption{\label{fig:gofr2D} Logarithm of the total correlation function $h_{bb}(r)$ for a two-dimensional binary mixture of  hard disks, with size ratio $q=0.5$, obtained from DFT (a) and MC simulations (b). $\eta_b=\rho_b \pi \sigma_b^2 / 4$ is fixed at 0.3 in all cases. For the packing fraction of the small disks $\eta_s=0.075$, the oscillations have a wavelength of about $\sigma_b$ whereas for $\eta_s=0.30$ the wavelength is about $\sigma_s = \sigma_b/2$. Interference effects are visible for intermediate values, especially for $\eta_s=0.1975$ in the DFT results (a), providing clear indication of structural crossover. The results are shifted vertically for clarity of display.}
\end{figure}

In order to study the crossover behaviour of correlation functions in $d=2$ we performed DFT calculations and standard MC simulations in the canonical ensemble for a binary mixture of hard disks with a size ratio of $q=0.5$. Within DFT we used the test-particle approach, fixing a single big disk at the origin and calculating the density profiles $\rho_i(r)$ of both components, $i=s,b$, in the external field exerted by the fixed disk. From the density profiles, we determined the correlation functions $h_{bi}(r)=\rho_i(r)/\rho_i - 1$.
This is the same procedure as in paper I, except now we employ Rosenfeld's $d=2$ 'interpolation' form of the free energy functional \cite{bib:rosenfeld90}. Throughout the calculation and simulation we kept the packing fraction of the big disks constant at $\eta_b=\rho_b \pi \sigma_b^2 / 4 =0.3$. We increased the packing fraction of the small disks from a very small value, $\eta_s=0.075$, to a high value, $\eta_s=0.3$. The resulting total correlation functions $h_{bb}(r)$ obtained from DFT, (a), and MC simulations, (b) are shown in figure \ref{fig:gofr2D}. The overall agreement between these two sets of results is encouraging. We find for the lowest value of $\eta_s$ that the wavelength of the decay of $h_{bb}(r)$ is set by $\sigma_b$, the diameter of the big disks, and for the highest value of $\eta_s$ that the wavelength is set by $\sigma_s$, the diameter of the small disks.  The DFT results for intermediate values of $\eta_s$ show evidence of interference and indicate clearly that crossover occurs near $\eta_s=0.1975$. The MC results show complex interference effects at the intermediate values of $\eta_s$. Although we did not determine the precise location of the crossover line in $d=2$, as we did in $d=1$ and $d=3$, our results demonstrate that structural crossover occurs in $d=2$.

The fact that crossover is reflected in the oscillatory behaviour for intermediate separation and can be seen in MC data, which contain some statistical noise, suggests that video microscopy in $d=2$ should be well suited to study various aspects of the decay of correlation functions in colloidal mixtures.

%%%%%%%%%%%%%%%%%%%%%%%%%%%%%%%%%%%%%%%%%%%%%%%%%%%%%%%%%%%%%%%%%%%%%%%%%%%%%%%%
%%%%%%%%%%%%%%%%%%%%%%%%%%%%%%%%%%%%%%%%%%%%%%%%%%%%%%%%%%%%%%%%%%%%%%%%%%%%%%%%
%%%%%%%%%%%%%%%%%%%%%%%%%%%%%%%%%%%%%%%%%%%%%%%%%%%%%%%%%%%%%%%%%%%%%%%%%%%%%%%%
\section{Discussion} \label{sec:discussion}

We have investigated the nature of structural crossover in various properties of binary hard-sphere mixtures. Whereas in paper I we focused only on the decay of pair correlation functions, $h_{ij}(r)$, in homogeneous (bulk) mixtures, here we consider manifestations of crossover for the one-body density profiles, $\rho_i(z)$, of the mixture adsorbed at planar walls for the solvation force, $f_s(L)$, and for the depletion potential, $W(r)$, of the same mixture. In all of these quantities we find a clear signature of structural crossover, i.e.~the wavelength of the oscillatory decay changes from being approximately the diameter of the big species to approximately that of the smaller species at some crossover point in the ($\eta_b$,$\eta_s$) phase diagram. We argue that the precise crossover point must be the same as that obtained from analysing the leading order poles of $\hat h_{ij}(k)$, as described in I. Moreover, we show that leading order asymptotics, as described by a single pole approximation such as equations (\ref{eqn:decaywall2}), (\ref{eqn:deplpotdecay}) and (\ref{eqn:sfdecay}) provides a remarkably good description of oscillatory structure at intermediate range as well as at longest range --- provided one is not too close to the crossover point, where a two-pole approximation such as (\ref{eqn:decaywallcrossover}) is required. This conclusion, which is equivalent to that ascertained in I for $h_{ij}(r)$, is important for practical purposes. Experiments or computer simulations are necessarily restricted to small or intermediate length scales --- one cannot probe the ultimate decay of structural properties. For example, if one were seeking evidence for structural crossover in measurements of the solvation force for an asymmetric binary mixture confined between the two mica plates of the surface force apparatus \cite{bib:israelachivili91} one would have to probe plate separations on the scale of a few particle diameters --- as in figure \ref{fig:solvationforce_log}. Similar remarks pertain to experimental measurements \cite{bib:deplpotmeasurements} of the depletion potential between two very big colloidal particles immersed in a binary mixture of big and small (hard-sphere) colloids. Crossover is already apparent at intermediate length scales --- see figure \ref{fig:depletion}. 

It is important to enquire whether the results we obtain here for the idealised system of hard spheres are relevant to real fluids, which could be atomic mixtures or mixtures of two types of colloidal particles. This issue was addressed in I for bulk mixtures and the remarks are equally pertinent here. Assuming that dispersion forces (power-law decay) play a minor role, which could be engineered by suitable refractive index matching in colloidal fluids, structural crossover should occur for mixtures where the effective pair potentials are softer than hard-spheres, provided the effective diameters are sufficiently different. Indeed the first observation of such crossover was found \cite{bib:archer01} in the binary Gaussian core mixture where the pair potentials are very soft. Of course, the presence of attractive interactions might force the regime where crossover effects can be observed to larger distances. This is the case for the binary mixtures adsorbed at planar walls that exhibit attractive wall-fluid potentials --- see figure \ref{fig:rho_attr_log}.

Our results for a simple model of polydisperse hard spheres adsorbed at a planar hard wall warrant further comment. We argue, in section \ref{sec:polydispersity}, that for a size distribution with two broad, weakly overlapping maxima, structural crossover remains visible in the density profile of the species $b$ corresponding to the maximum packing fraction of the bigger spheres --- see figure \ref{fig:rhowallpoly}. At first sight this might appear surprising. Consider first a distribution of big hard-sphere diameters with a single maximum at $\sigma_b^{max}$. Provided the number of diameters is finite, corresponding to a finite number of species $\nu$, the OZ equations can still be written in the form of equation (\ref{eqn:FourierOZ}) but now generalised to $\nu$ components. Nevertheless, $\hat h_{ij}(k)$, with $i,j = 1 \ldots \nu$, will have poles determined by the zeros of the appropriate $\nu$ component common denominator $\hat D(k)$. There will be a LOP and the quantities $a_0$ and $a_1$ will be uniquely determined. For a narrow distribution these quantities will be close to those for a one-component fluid with diameter $\sigma_b^{max}$ at the same (total) packing fraction. Similarly for a fluid with a distribution of $\nu$ small hard-sphere diameters, with a narrow peak at $\sigma_s^{max}$, we expect the LOP to be close to that for the one-component fluid with diameter $\sigma_s^{max}$. When we consider the polydisperse 2$\nu$ component mixture, corresponding to the bimodal situation of figure~\ref{fig:distribution}, equation (\ref{eqn:FourierOZ}) remains valid and we should expect to find that the LOPs, determined by the zeros of the 2$\nu$ denominator, are not very different from those in the binary mixture which corresponds to the size ratio $\sigma_s^{max}/\sigma_b^{max}$. Although $\nu$ might be large (we consider $\nu$ = 21 in section \ref{sec:polydispersity}), provided it is finite and the ratio $\sigma_s^{max}/\sigma_b^{max}$ remains sufficiently small, we do expect to find crossover. Of course, as described in section \ref{sec:polydispersity}, the crossover need not occur at the same point in the phase diagram as in the binary mixture. 

We have not attempted to incorporate genuine polydispersity, i.e.~a continuous distribution of diameters corresponding to an infinite number of species, into our theoretical treatment. We leave this for future study, noting that detailed DFT and Monte Carlo studies \cite{bib:pagonabarra00,bib:buzzacchi04} have investigated the density profiles of polydisperse hard-spheres at a planar hard-wall. However, these studies employ a parent bulk distribution (top-hat or Schulz) that has a single maximum and are not concerned with the intermediate range oscillatory behaviour of the profiles. In experimental samples, typical degrees of size-polydispersity are around 5\%, which is less than the degree of polydispersity considered here. Therefore we are confident that the structural crossover can be observed in experiments with colloidal mixtures.

Finally we return to our results for one-dimensional hard-rods (subsection \ref{sec:onedim}). It is gratifying that structural crossover occurs in an exact statistical mechanical treatment and that the overall shape of the crossover line follows that calculated for the binary mixture of hard spheres. This attests to the generic nature of the crossover mechanism. In both one and three dimensional fluids with $q = 0.5$ the line begins at the origin of the ($\eta_b$,$\eta_s$) plane. (Note that our results in figure 5 of I did not extend to very low packing fractions.) For a very low concentration of small particles crossover occurs rapidly on adding a small amount of big particles. The form of the crossover line in the limit $\eta_b \to 1$ for the hard-rod mixture in figure \ref{fig:crossoverline1d} is somewhat more curious. If the crossover line exists in this limit it must, of course, terminate at the point $\eta_b = 1$, $\eta_s = 0$ since the total packing fraction cannot exceed unity. Nevertheless, it is striking that for a very large concentration of big rods, say $\eta_b = 0.97$, adding a very small amount of small particles so that the total packing fraction is about 0.98, already leads to crossover, i.e.~the wavelength of the longest range of oscillations in the correlation functions is then determined by the length of the small rods which are present in only very small concentrations. We should note, however, that as $\eta_b \to 1$ several poles begin to contribute to $h_{ij}(r)$ at long-range; the imaginary parts of the higher-order poles become comparable with $a_0$, the imaginary part of the LOP \cite{bib:comment1d}.

Michael Fisher and Ben Widom \cite{bib:fisher69} pioneered investigations of how the character of the asymptotic decay of the pair correlation function should change as one moves from one region of the phase diagram to another. Their study is relevant to fluids in which there are both repulsive and attractive contributions to the interparticle potential and the existence of the FW line is intimately linked to the presence of liquid-gas, or in the case of mixtures, fluid-fluid phase separation. The structural crossover we describe here and in I arises in fluid mixtures where the interparticle potentials are purely repulsive; what is required is that the two species are sufficiently different in size. As mentioned in the introduction, FW based their study on exactly solvable one-dimensional models. For reasons that will be abundantly clear to the reader, we are grateful that they did not consider mixtures! 

\ \\
R.E.~thanks S.~Dietrich and his colleagues for their hospitality during his visits to the MPI in Stuttgart.

%%%%%%%%%%%%%%%%%%%%%%%%%%%%%%%%%%%%%%%%%%%%%%%%%%%%%%%%%%%%%%%%%%%%%%%%%%%%%%%%
%%% bibliography %%%%%%%%%%%%%%%%%%%%%%%%%%%%%%%%%%%%%%%%%%%%%%%%%%%%%%%%%%%%%%%
%%%%%%%%%%%%%%%%%%%%%%%%%%%%%%%%%%%%%%%%%%%%%%%%%%%%%%%%%%%%%%%%%%%%%%%%%%%%%%%%

\end{document}